\documentclass[journal,10pt]{IEEEtran}

\usepackage{cite}
\usepackage{graphicx}
\usepackage{amsmath}
\usepackage{array}
\usepackage{mdwmath}
\usepackage{amssymb}
\usepackage{mdwtab}
\usepackage{stfloats}
\usepackage[tight,footnotesize]{subfigure}
\usepackage{amsmath,amsthm}
\usepackage{threeparttable}
\usepackage{color}
\usepackage{url}
\usepackage{algpseudocode}
\usepackage{algorithm}
\usepackage{multicol}
\usepackage{amssymb}
\usepackage{epstopdf}

\algrenewcommand{\algorithmicrequire}{\textbf{Input:}}
\algrenewcommand{\algorithmicensure}{\textbf{Output:}}

\newtheorem{proposition}{Proposition}

\newtheorem{theorem}{Theorem}

\newtheorem{lemma}{Lemma}

\ifCLASSINFOpdf
\else
\fi

\begin{document}

\title{RIS-Aided Spatial Scattering Modulation for mmWave MIMO Transmissions}

\author{
Xusheng Zhu, Wen Chen, \IEEEmembership{Senior Member, IEEE}, Zhendong Li, Qingqing Wu, \IEEEmembership{Senior Member, IEEE}, \\Ziheng Zhang, Kunlun Wang, \IEEEmembership{Member, IEEE}, and Jun Li, \IEEEmembership{Senior Member, IEEE}
\thanks{X. Zhu, W. Chen, Z. Li, Q. Wu, and Z. Zhang are with the Department of Electronic Engineering, Shanghai Jiao Tong University, Shanghai 200240, China (e-mail: xushengzhu@sjtu.edu.cn; wenchen@sjtu.edu.cn; lizhendong@sjtu.edu.cn; qingqingwu@sjtu.edu.cn; zhangziheng@sjtu.edu.cn).}
\thanks{K. Wang is with the School of Communication and Electronic Engineering, East China Normal University, Shanghai 200241, China (e-mail:
klwang@cee.ecnu.edu.cn).}
\thanks{J. Li is with the School of Electronic and Optical Engineering,
Nanjing University of Science Technology, Nanjing 210094, China (e-mail:
jun.li@njust.edu.cn).}
}

\markboth{}
{}
\maketitle
\begin{abstract}
This paper investigates the reconfigurable intelligent surface (RIS) assisted spatial scattering modulation (SSM) scheme for millimeter-wave (mmWave)  multiple-input multiple-output
(MIMO) systems, in which line-of-sight (LoS) and non-line-of-sight (NLoS) paths are respectively considered in the transmitter-RIS and RIS-receiver channels. Based on maximum likelihood detector, the conditional pairwise error probability (CPEP) expression for the RIS-SSM scheme is derived
under the two cases of received beam correct and demodulation error.
Furthermore, we derive the closed-form expressions of the unconditional pairwise error probability (UPEP) by employing two different methods: the probability density function and the moment-generating function expressions with a descending order of scatterer gains.
To provide more useful insights, we derive the asymptotic UPEP and the diversity gain of the RIS-SSM scheme in the high SNR region. Depending on UPEP and the corresponding Euclidean distance, we get the union upper bound of the average bit error probability (ABEP). To acquire the effective capacity of the proposed system, a new framework for ergodic capacity analysis is also provided. Finally, all derivation results are validated via extensive Monte Carlo simulations and reveal that the proposed RIS-SSM scheme outperforms the benchmarks in terms of reliability.
\end{abstract}


\section{Introduction}
In the past decade, there has been a significant increase in demand for higher data rates and the rapid expansion of data services, leading to the emergence of a plethora of innovative wireless communication technologies. One such technology is massive multiple-input multiple-output (MIMO), which has shown significant potential in improving the quality of communication by increasing the number of antennas at either the transmitter (Tx) or the receiver (Rx) \cite{lu2014an}. However, the cost of equipping each antenna with an expansive radio frequency (RF) chain has led to high power consumption and cost, which has resulted in the need for a completely novel communication paradigm to address these challenges in the physical layer \cite{easar2019wirel}.

Recently, reconfigurable intelligent surface (RIS) has emerged as a highly promising technique for the development of 6G wireless networks. This is due to their unique ability to effectively alter the propagation environment while simultaneously reducing power consumption and hardware cost. Specifically, RIS is comprised of a large number of low-cost reflective elements arranged in a two-dimensional (2D) planar array \cite{pan2021reconfi,zhu2023per}. Each element of the RIS can be regarded as a reconfigurable scatter, capable of reflecting the incident signal by tuning the phase shift \cite{wu2020intelligent,yang2020acc}. By jointly manipulating all elements of the RIS, the reflection signal can be constructively enlarged to improve the signal-to-noise ratio (SNR) or degraded to mitigate interference \cite{li2022joint}. Additionally, RIS has an easy deployment feature, which can improve the delivery of communication services in the desired direction and extend the network coverage \cite{yang2021out,pan2021reconfigurable}. Compared with traditional relays, RIS does not require RF chains and amplifiers, which can greatly reduce communication energy consumption \cite{wu2019intelligent}.
It is worth noting that the authors of \cite{li2022beam} conducted a study on the transmissive RIS-based Tx architecture, while the author of \cite{mu2022simu} investigated the simultaneously transmitting and reflecting (STAR) RIS scheme by combining the transmissive and reflective RISs. These advancements in RIS technology make it a promising solution to the challenges of energy consumption and hardware cost in the next-generation communication networks.

Index modulation (IM) is a well-established concept in wireless communication that has gained considerable attention in recent decades. IM can facilitate the transmission of additional data by utilizing available resources such as transmit antenna, subcarrier, etc., \cite{mao2019nov}.
One of the prominent IM techniques is spatial modulation (SM), which utilizes one RF chain to establish connection with an arbitrary antenna, allowing the spatial location of the antenna to convey additional bits \cite{mes2008spati}.
It is worth noting that SM strikes a favorable balance between energy efficiency and spectral efficiency \cite{renzo2011spatial}.
A recent large-scale measurement campaign was conducted in various indoor environments to investigate the propagation characteristics of SM in real-world scenarios \cite{zhu2022on}.
Moreover, the variant of SM is quadrature SM (QSM), which was proposed to improve the overall throughput of conventional SM systems by incorporating additional modulation spatial dimensions \cite{mesleh2015qua}.
Space shift keying (SSK) is a simplified version of SM that relies only on the antenna index to achieve information transmission \cite{jeg2009spac}.
However, in high-frequency bands such as {millimeter-wave (mmWave)}, it is challenging to ensure the signal quality by activating only one antenna. To address this issue, spatial scattering modulation (SSM) was introduced in \cite{ding2017spatial}, which utilizes analog and hybrid beamforming structures to convey extra bits via the spatial angle domain.
Similar to SM, SSM employs a single RF chain, which allows the Tx to focus the beam in one direction at each time slot based on the input bits. Moreover, quadrature SSM (QSSM) scheme was also investigated in \cite{tu2018iccc}. Note that QSSM scheme aims to improve the performance of SSM by incorporating adaptive and generalized modulation schemes.

Motivated by the advantages of RIS and IM, researchers have explored the idea of incorporating IRS into RIS-based systems. This concept has been widely investigated in recent studies  \cite{easar2020rec,ma2020large,can2020rec,can2022on,li2021space,bho2021ris,din2022ris,zhu2022re},
which can be categorized into three groups: RIS-SM, RIS-SSK, and RIS-SSM schemes.
Specifically, for the RIS-SM schemes, Basar proposed employing SM on receive antenna indices. Here, the RIS aims to transmit a beam at a specific receive antenna to enhance the spectral efficiency based on the spatial bit information \cite{easar2020rec}.
Further, the authors of \cite{ma2020large} deploys RIS to the middle of the channel, and simultaneously introduces transmit antenna index and receive antenna index information into RIS-SM, resulting in a significant increase in spectral efficiency.
Additionally, \cite{yang2021a} presented a RIS-assisted multiuser uplink communication system incorporating a novel modulation scheme.
For the RIS-SSK schemes, the authors of \cite{can2020rec} and \cite{wu2020intelligent} reported the ABEP performance over Rayleigh channels, where the SSK is realized at the Tx side.
Building on this work, \cite{can2022on} investigated power sensing and partitioned RIS-SSK schemes in order to enhance system performance.
In pursuit of both spectral-efficient and energy-efficient communications, the authors of \cite{bho2021ris} investigated RIS-aided full-duplex SSK systems.
To achieve reliable transmission, \cite{li2021space} proposed RIS-SSK schemes with passive beamforming and RIS-SSK Alamouti space-time block coding.
Additionally, the authors of \cite{din2022ris} presented RIS-assisted receive quadrature SSK (RIS-RQSSK), which decomposes the spatial domain to extend the spectral efficiency. In this scheme, the goal of the RIS is to maximize the SNR of the in-phase and quadrature components.
For the RIS-SSM scheme, This scheme was proposed in \cite{zhu2022re} as an alternative to RIS-SM/SSK schemes in the mmWave band. The RIS-SSM scheme involves replacing the activated antenna in RIS-SM/SSK with a beam directed towards the scatterer, which is promising for mmWave communication to increase spectral efficiency.
{
It is worth mentioning that the proposed RIS-SSM scheme has a potential application in secure communications. Specifically, the codes represented by the participating scatterers is known only to the Tx and Rx.
Meanwhile, the eavesdropper is hard to identify the code by the participated scatterers, especially in mobile communication scenarios where the scatterers are changing time by time.
}

\begin{figure*}[t]
\centering
\includegraphics[width=17.5cm]{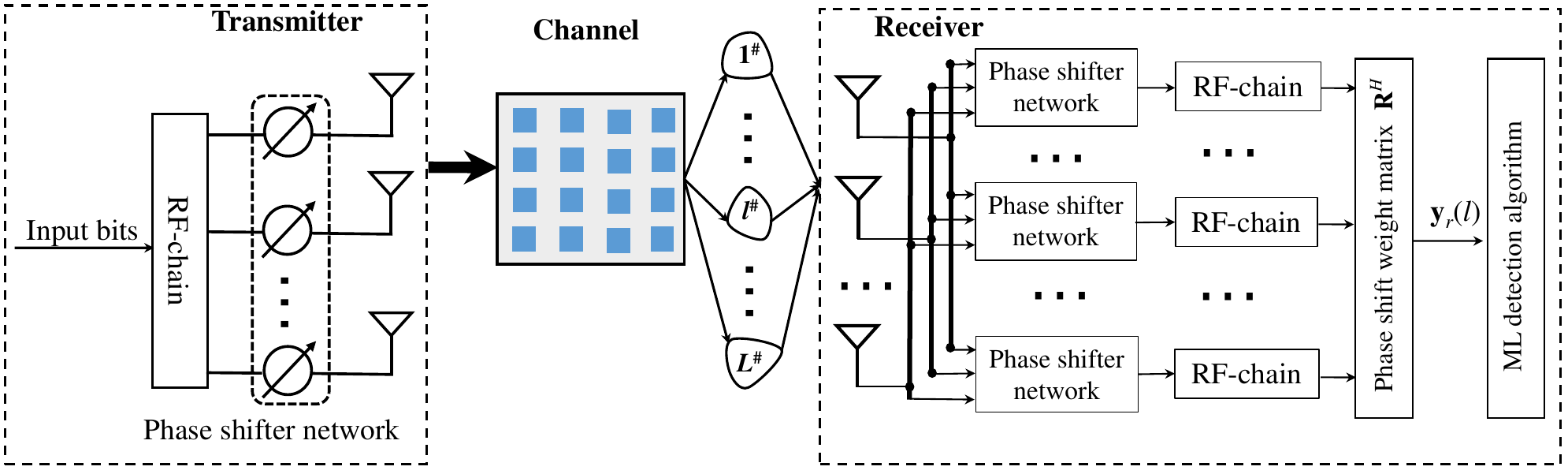}
\caption{\small{System model of the proposed RIS-SSM scheme.}}
\vspace{-10pt}
\label{sys}
\end{figure*}
In light of the aforementioned considerations, we propose a novel RIS-SSM scheme, aimed at achieving energy-efficient wireless communication. While the first work on RIS-assisted SSM was proposed in \cite{zhu2022re}, the present study introduces two key differences between the two {respects}: 1) System model: This paper adopts a line-of-sight (LoS) path communication in the Tx-RIS channel, while the SSM is utilized in the RIS-Rx channel. In contrast, both Tx-RIS and RIS-Rx channels employ SSM under none-line-of-sight (NLoS) paths to facilitate information transmission in \cite{zhu2022re}. 2) Analytical derivations: The current work derives a series of closed-form analytical expressions and investigates the ergodic capacity and system throughput of the proposed scheme. In contrast, the authors of \cite{zhu2022re} fail to provide a closed-form analytical average bit error probability (ABEP) expression.
To the best of our knowledge, no systematic analysis of the performance of RIS-assisted SSM schemes has been conducted in the existing literature. In order to obtain a comprehensive understanding of the potential of this approach in wireless communication networks,
the main contributions of this work are summarized as follows:
\begin{itemize}
\item[1)] In this paper, we consider a RIS-assisted SSM mmWave transmission system, where the Tx and RIS are deployed relatively closed positions, they can communicate through the LoS path, while the distance between the RIS and Rx is relatively far from each other, the NLoS paths are considered to convey via the SSM technique is employed in the RIS-Rx channel.
    It is worth mentioning that the scatterer amplitude gain of the proposed RIS-SSM scheme is in descending order. At the Rx side, we adopt the maximum likelihood (ML) detector for the joint demodulation of the spatial domain beam and the symbol domain signal.
\item[2)] A detailed derivation of the conditional pair error probability (CPEP) for RIS-SSM is performed based on ML  detection algorithm. Then, we consider both correct and wrong beam detection scenarios. Subsequently, the probability density function (PDF) and moment generating function (MGF) expressions based on the magnitude gain of the scatterers arranged in descending order are derived, respectively. Furthermore, we obtain a closed-form expression for the unconditional pair error probability (UPEP) using the derived PDF and MGF.  In addition, a closed-form expression for UPEP is provided based on the traditional Q-function estimation method.
\item[3)] {To gain more insights, we derived the ABEP asymptotically closed-form expression of the UPEP and the diversity gain of the RIS-SSM system. Moreover, in order to effectively characterize the capacity of the proposed scheme, we also derive the lower bound of the ergodic capacity. Meanwhile, the lower bound expression of the system throughput of the proposed RIS-SSM scheme is provided in this work.}
\item[4)] Monte Carlo simulations are conducted in this paper to validate all the analytical results and demonstrate the following findings: Firstly, the proposed RIS-SSM scheme achieves significantly better performance than the maximum and minimum beam schemes when the channel environment has a higher density of scatterers. Secondly, the modulation scheme based on descending order of scattering gain outperforms the random arrangement scheme. Thirdly, the two ABEP derivation methods proposed in this paper outperform the approach based on Q-function estimation method. Lastly, the richness of scatterers has a positive impact on the ABEP performance and the ergodic capacity improves, as the number of scatterers in the channel increases.
\end{itemize}

The rest of this paper is organized as follows. System model of the RIS-SSM scheme is described  in Section II.  Section III provides a comprehensive mathematical analysis of the ABEP, capacity, and throughput of the proposed RIS-SSM scheme. In the Section IV, the important results and behaviors are presented and discussed. Finally, Section V gives a summary of this paper.

\emph{Notations:}
Lowercase and uppercase letters are used to denote the column vectors and matrices, respectively.
$(\cdot)^T$ and $(\cdot)^H$ denote the transpose and complex conjugate transpose operations.
$\mathbf{I}_n$ denotes the $n\times n$ identity matrix.
${\rm diag}(\cdot)$ represents diagonal matrix operation.
$\mathbb{C}^{n \times m}$ is the space of ${n \times m}$ complex-valued matrices.
The real part of a complex variable is represented by $\Re\{\cdot\}$.
$\Pr(\cdot)$ means the probability of an event, while $\exp(\cdot), Q(\cdot),\mathcal{B}(\cdot,\cdot)$, and  $\Gamma(\cdot)$ denote Exponential function, Q-function, Beta function, and Gamma function, respectively.
The modified bessel function of the first kind of zeroth order is indicated as $I_0(\cdot)$.
Moreover, $\delta(\cdot)$ denotes Dirac function.
Factorial operation denotes $(\cdot)!$, while
$(\cdot)!!$ represents double factorial operation.
$\|\cdot\|$ is Frobenius-norm operation.
$f(\cdot)$ is the PDF in descending order.
$\ddot f(\cdot)$ denotes the PDF without the statistical order calculation method.
$\mathcal{N}(0,\sigma^2)$ represents real Gaussian distribution with covariance $\sigma^2$, while
$\mathcal{CN}(0,\sigma^2)$ is used to denote circularly symmetric complex Gaussian
distribution with covariance $\sigma^2$.
$\mathbb{E}(\cdot)$ stands for statistical expectation operation.
PSK/QAM denotes phase shift keying/quadrature amplitude modulation.

\section{System Model}
In Fig. \ref{sys}, we consider a RIS-assisted narrowband hybrid mmWave MIMO transmission system model,
where the RIS is deployed in the channel to assist the information transmission.
It is worth noting that we assume that the RIS is deployed close to the Tx and can communicate directly with an LoS path.
On the other hand, the RIS is deployed far from the Rx.
In this case, we cannot guarantee that an LoS path still exists between RIS and Rx.
Hence, it is reasonable to model the communication between RIS and Rx as a NLoS-dominated channel \cite{ding2017spatial}.

\subsection{RIS-SSM Channel Model}
In this paper, we assume that Tx and Rx are equipped with uniform linear array (ULA), while the RIS is composed of the 2D uniform planar array (UPA).
\begin{figure*}[t]
\begin{minipage}[t]{0.45\linewidth}
\centering
\includegraphics[width=7cm]{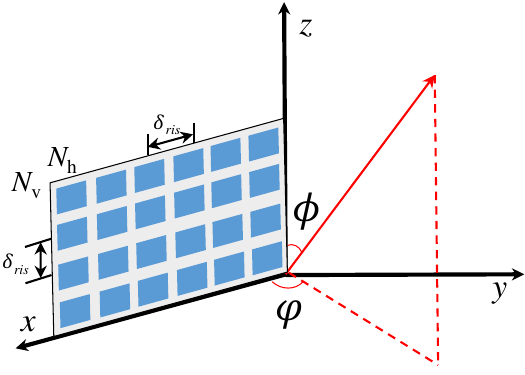}
\caption{\small{3D coordinate diagram of RIS.}}
\label{Figris}
\end{minipage}%
\hfill
\begin{minipage}[t]{0.45\linewidth}
\centering
\includegraphics[width=7cm]{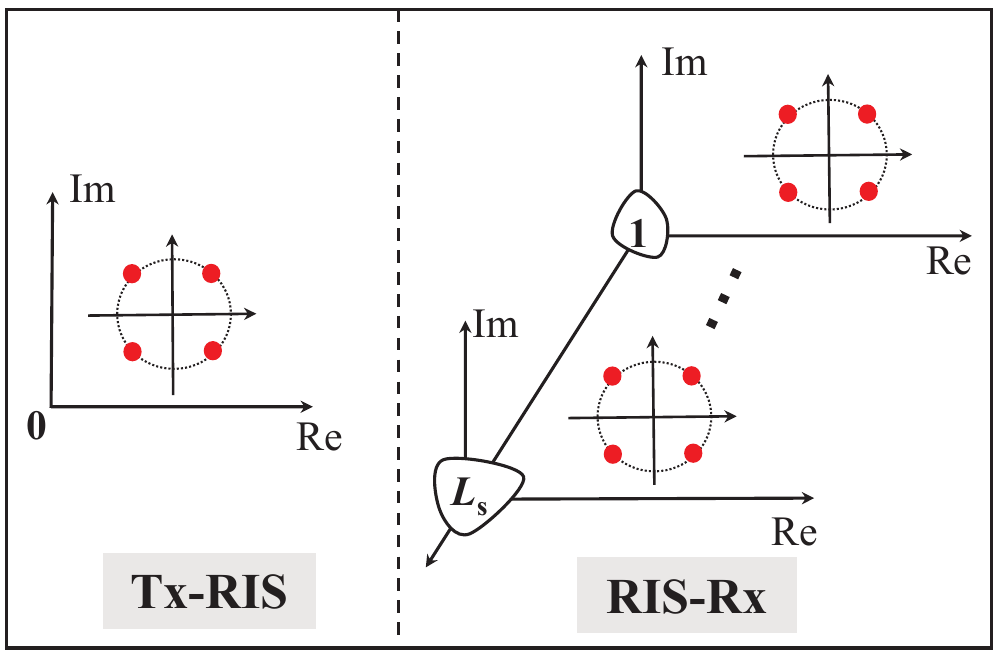}
\caption{\small{Constellation with QPSK and $L_s$. }}
\label{diagratt}
\end{minipage}
\end{figure*}
\subsubsection{Tx-RIS Channel}
In this situation, we assume that there exists the strong LoS mmWave MIMO link. Thus, the channel between the Tx and RIS can be characterized as
\begin{equation}\label{B1}
\mathbf{B} = \boldsymbol{a}_{ ris}(\phi^r,\varphi^r)\boldsymbol{a}_{t}^H(\theta^t),
\end{equation}
where $\boldsymbol{a}_{t}(\theta^t)$ represents the response of the $N_t$-element ULA at the Tx.
According to \cite{ding2017spatial}, $\boldsymbol{a}_{t}(\theta^t)$ can be written as
\begin{equation}
\boldsymbol{a}_{t}(\theta^t)=\frac{1}{\sqrt{N_t}}
\left[
1,e^{-j2\pi\frac{\delta_t}{\lambda}\sin\theta^t},\cdots,
e^{-j2\pi\frac{\delta_t}{\lambda}(N_t-1)\sin\theta^t}
\right]^T,
\end{equation}
where $\delta_t$ represents the spacing between adjacent elements in the Tx array, $\lambda$ means the carrier wavelength, $\theta^t \in\left[-{\pi},{\pi}\right]$ represents the angle of departure (AoD) of signals from the Tx.
As shown in Fig. \ref{Figris}, $\boldsymbol{a}_{ ris}^H(\phi^r,\varphi^r)$ stands for the array response at the RIS in the Tx-RIS channel, which can be evaluated as
\begin{equation}\label{upar}
\begin{aligned}
&\boldsymbol{a}_{ ris}(\phi^r,\varphi^r) = \frac{1}{\sqrt{N}}
\left[
1, \cdots,e^{-j2\pi\frac{\delta_{ris}}{\lambda}(n_{\rm h}\sin\phi^r\cos\varphi^r+
n_{\rm v}\cos\phi^r)},\right.\\&\left.\cdots,e^{-j2\pi\frac{\delta_{ris}}{\lambda}((N_{\rm h}-1)\sin\phi^r\cos\varphi^r+(N_{\rm v}-1)\cos\phi^r)}
\right]^T,
\end{aligned}
\end{equation}
where $N_{\rm h}$ and $N_{\rm v}$ are the number of rows and columns of RIS, respectively.
At the same time, $n_{\rm h}$ and $n_{\rm v}$ represent the corresponding row index and column index, respectively.
$\delta_{ris}$ denotes the spacing between adjacent reflective RIS elements.
$\phi^r \in \left[-{\pi},{\pi}\right]$ represents the elevation angle of arrival (AoA) of signals to the RIS and $\varphi^r \in \left[-{\pi},{\pi}\right]$ denotes the azimuth AoA of signals to the RIS.

\subsubsection{RIS-Rx Channel}
{
Due to the smaller wavelength of millimeter waves, their diffraction capability is limited when encountering obstacles. As a result, the millimeter wave channel exhibits a sparse multipath structure, often characterized by the Saleh-Valenzuela (SV) channel model. It is worth noting that the SSM technique is employed to achieve information transmission in the RIS-Rx channel operating in the millimeter wave frequency range. Therefore, the RIS-Rx channel can be represented as follows:
}
\begin{equation}\label{F}
\mathbf{F}= \sum_{l=1}^Lh_l \boldsymbol{a}_r(\theta_{l}^r)
\boldsymbol{a}_{ris}^H(\phi_l^t,\varphi_l^t),
\end{equation}
where $L$ is the total number of signal paths between the RIS and the Rx.
$\phi_l^t \in \left[-{\pi},{\pi}\right]$ and $\varphi_l^t \in \left[-{\pi},{\pi}\right]$ denote the azimuth and elevation AoDs associated with the RIS, respectively. $\theta_l^r \in \left[-{\pi},{\pi}\right]$ represents the AoA at the Rx side, $h_l$ is the complex channel gain.
$\boldsymbol{a}_{r}(\theta^r)$ denotes the normalized array response vectors associated with the Rx.
Moreover, the ULA with $N_r$ antennas at the Rx side can be given as
\begin{equation}
\boldsymbol{a}_{r}(\theta_l^r)=\frac{1}{\sqrt{N_r}}
\left[
1,e^{-j2\pi\frac{\delta_r}{\lambda}\sin\theta_l^r},\cdots,
e^{-j2\pi\frac{\delta_r}{\lambda}(N_r-1)\sin\theta_l^r}
\right]^T,
\end{equation}
where $\delta_r$ and $\lambda$ are the antenna spacing and the signal wavelength.
Besides, the normalized array response transmitted from the RIS can be given by
\begin{equation}\label{upat}
\begin{aligned}
&\boldsymbol{a}_{ ris}(\phi_l^t,\varphi_l^t) =\frac{1}{\sqrt{N}}
\left[
1, \cdots,e^{-j2\pi\frac{\delta_{ris}}{\lambda}(n_{\rm h}\sin\phi_l^t\cos\varphi_l^t+n_{\rm v}\cos\phi_l^t)},\right.\\&\left.\cdots,
e^{-j2\pi\frac{\delta_{ris}}{\lambda}((N_{\rm h}-1)\sin\phi_l^t\cos\varphi_l^t+(N_{\rm v}-1)\cos\phi_l^t)}
\right]^T.
\end{aligned}
\end{equation}

\subsubsection{Composite Channel }
The transmitted signal propagates through the Tx-RIS-Rx channel to reach Rx, where the RIS consists of $N$ passive reflecting elements.
Each of them is independent of each other without interference.
To characterize the performance limit of RIS, we assume the magnitude of each reflecting element equals one.
As such, the reflection matrix of the RIS can be formulated as
\begin{equation}\label{psi}
\boldsymbol{\Psi}={\rm diag}(e^{j\psi_1},\cdots,e^{j\psi_n},\cdots,e^{j\psi_N}) \in \mathbb{C}^{N\times N}.
\end{equation}
Considering Eqs. (\ref{B1}), (\ref{F}), and (\ref{psi}), the composite channel can be calculated  as
\begin{equation}\label{h11}
\begin{aligned}
\mathbf{H}&= \mathbf{F} \boldsymbol{\Psi} \mathbf{B}= \sum_{l=1}^Lh_l \boldsymbol{a}_r(\theta_{l}^r)
\boldsymbol{a}_{ris}^H(\phi_l^t,\varphi_l^t) \boldsymbol{\Psi}
\boldsymbol{a}_{ ris}(\phi^r,\varphi^r)\boldsymbol{a}_{t}^H(\theta^t).
\end{aligned}
\end{equation}
We define $\beta_l\overset{\Delta}{=}h_l\boldsymbol{a}_{ris}^H(\phi_l^t,\varphi_l^t) \boldsymbol{\Psi}
\boldsymbol{a}_{ ris}(\phi^r,\varphi^r)$. Then, the Eq. (\ref{h11}) can be recast as
\begin{equation}\label{h001}
\begin{aligned}
\mathbf{H}
&= \sum_{l=1}^L\beta_l \boldsymbol{a}_r(\theta_{l}^r)
\boldsymbol{a}_{t}^H(\theta^t).
\end{aligned}
\end{equation}
Herein, we set
$\zeta^r = \arg [\boldsymbol{a}_{ ris}^H(\phi^r,\varphi^r)]$, and
$\zeta^t_l = \arg [\boldsymbol{a}_{ris}(\phi_l^t,\varphi_l^t)]$.
Combing $\zeta^r$, $\zeta^t_l$, and Eq. $(\ref{psi})$, the phase ${\Xi}$ of composite channel can be described as
$
{\Xi}
= \frac{1}{{N}}\sum_{n=1}^Ne^{j(\psi_n-\zeta^t_l-\zeta^r)}.
$
By tuning the phase shifts of RIS as $\psi_n=\zeta^t_l+\zeta^r$, we have $\sum_{n=1}^Ne^{j(\psi_n-\zeta^t_l-\zeta^r)} = N$.
Hence, the Eq. (\ref{h001}) can be rewritten as
\begin{equation}\label{eq12}
\begin{aligned}
\mathbf{H}
&= \sum_{l=1}^Lh_l \boldsymbol{a}_r(\theta_{l}^r)
\boldsymbol{a}_{t}^H(\theta^t).
\end{aligned}
\end{equation}

\begin{lemma}
As the number of antennas of $N_t$ and $N_r$ is large enough, we can argue that the beams directed between the different scatterers are orthogonal to each other, which can be expressed as
\begin{equation}\label{ortho}
\begin{aligned}
\boldsymbol{a}_t^H(\theta_l^t)\boldsymbol{a}_t(\theta_{l'}^t) = \delta(l-l'), \ \ \boldsymbol{a}_r^H(\theta_l^r)\boldsymbol{a}_r(\theta_{l'}^r) = \delta(l-l').
\end{aligned}
\end{equation}
\end{lemma}

{\emph{Proof:}}
 Please refer Appendix A for the detailed proof. $\hfill\blacksquare$

\subsection{RIS-SSM Transmission}
In Fig. \ref{sys}, we consider that the RIS is deployed close to the Tx, where the Tx and RIS communicate through the LoS link. Among the RIS-Rx channels, it is difficult to guarantee the existence of a stable LoS path due to the existence of abundant scatterers. Consequently, the RIS-Rx can realize communication through the NLoS path.
In the proposed RIS-SSM scheme, the SSM technique is employed in the RIS-Rx channel, where the RIS can accurately emit the reflecting beam towards the objective scatterer in each time slot based on the perfect channel state information (CSI).
To enhance the reliability of the transmission, we select $L_s$ candidate scatterers in descending order of amplitude gain out of the $L$ scatterers in the RIS-Rx channel, i.e., $|h_1|>|h_2|>\cdots>|h_{L}|$.
Thus, the received information at the Rx side can be given as
\begin{equation}
\mathbf{y} = \sqrt{{P_s}}\mathbf{H}\boldsymbol{a}_{t}(\theta^t) s_m + {\mathbf{n}},
\end{equation}
where $P_s$ denotes the transmit power, and $\mathbf{n}$ represents additive Gaussian white noise vector following $\mathcal{CN}(0,N_0\mathbf{I}_{N_r})$.

In the proposed RIS-SSM, the bit stream can be divided into two parts, one is delivered by the scatterer and the other is carried by the conventional PSK/QAM signal.
To intuitively understand the modulation mechanism of RIS-SSM, we draw the corresponding constellation diagram in Fig. \ref{diagratt}.
In this figure, we utilize the QPSK in the conventional symbol domain and $L_s$ scatterers in the spatial domain to participate in the modulation.
It is obvious from Fig. \ref{diagratt} that only 2 bits per channel use (bpcu) signal is conveyed in the Tx-RIS channel, while $\log_2(2L_s)$ bpcu signal is conveyed in the RIS-Rx channel.
\subsection{RIS-SSM Detection}

The maximum ratio combing (MRC) technique is applied to achieve reception of the signal from the $l$-th scatterer at the Rx side.
Since the RIS transmit beam is aimed at a specific scatterer, interference between scatterers can be avoided entirely.
To distinguish that the received beam is coming from a specific scatterer, the Rx is equipped with ${R}_{r} \geq L_s$ RF chains.
Notice that each RF chain has a set of phase shifters to monitor a specific scatterer.
In this respect, it is feasible to monitor the $L_s$ scattered beams in real time, thus enabling the recovery of the signal in the spatial domain.
Here, the combined signal after the RF chains can be given by
\begin{equation}
\mathbf{y}_r=
[\boldsymbol{a}_r^H\left(\theta_1^r\right),\dots,\boldsymbol{a}_r^H\left(\theta_{L_{\mathrm{s}}}^r\right)]^T
\mathbf{y}.
\end{equation}
The ML detector is performed to jointly detects the activated scatterer index and the modulated symbols, which can be characterized as
\begin{equation}\label{ml}
{[{\hat{l}}, {\hat{m}}]}=
\mathop{\mathop{\mathop\mathbf{{\arg\min}} _{{m \in\{1, \ldots, M\}}}}_{{l \in\{1, \ldots, L_s\}}}}
|\mathbf{y}_{{r}}\left(l\right)-\boldsymbol{a}_r^H(\theta_{ l}^r)\mathbf{H}{\boldsymbol{a}_t(\theta^t)}\sqrt{P_s}s_m |^{2},
\end{equation}
where $\mathbf{y}_r(l)$ denotes the $l$-th element of $\mathbf{y}_r$,
$\hat l$ and $\hat s_m$ denotes the detected index of scatterer and symbol, respectively.

\section{Performance Analysis}
In this section, we first derive the union upper bound of CPEP expression by the ML detection algorithm.
Based on this, we give the UPEP expression of proposed scheme with two approaches.
Next, we provide the asymptotic UPEP and diversity gain of the proposed scheme.
Furthermore, we derive exact analytical expressions for ergodic capacity.
Finally, we provide the expression of system throughput.

\subsection{CPEP Expression}
To obtain the analytical ABEP of the RIS-SSM scheme, we first derive the CPEP for union detection of the selected scatterer index $l$ and the transmitted symbol $m$.
Specifically, mathematical derivation process can be divided into Rx correctly detected and falsely detected scatterers, where the detailed derivation is as follows:
\subsubsection{$\hat{l} = l$}
In this case, the detection of transmission direction is correct, whereas the error comes from the transmitted symbol, i.e., $s_m \neq s_{\hat m}$.
Herein, the CPEP can be calculated as
\begin{equation}\label{pr1}
\begin{aligned}
P_e=&{\Pr}\left(\left[l, {m}\right] \rightarrow[\hat{l}, \hat{m}] \right)\\
=&{\Pr}\left(|\mathbf{y}_r (l) - \boldsymbol{a}_r^H(\theta_{ l}^r)\mathbf{H}\boldsymbol{a}_t(\theta^t)\sqrt{P_s}s_m |^2\right.\\&\left.
>|\mathbf{y}_r ({ l}) - \boldsymbol{a}_r^H(\theta_{\hat l}^r)\mathbf{H}\boldsymbol{a}_t(\theta^t)\sqrt{P_s}s_{\hat m} |^2\right).
\end{aligned}
\end{equation}
Combining Eqs. (\ref{ortho}) and (\ref{pr1}), we can get
\begin{equation}\label{rorth}
\begin{cases}
\mathbf{y}_{{r}}\left(l\right)
= \mathbf{R}^H  \mathbf{y}\left(l,m\right)
=
\sqrt{P_s}h_{l}s_m+\boldsymbol{a}_r^H \left(\theta_{l}^r\right)\mathbf{n},\\
|\mathbf{y}_r ({l}) - \boldsymbol{a}_r^H(\theta_{ l}^r)\mathbf{H}{\boldsymbol{a}_t(\theta^t)}\sqrt{P_s}s_m |^2
=|\boldsymbol{a}_r^H(\theta_l^r)\mathbf{n}|^2,\\
\left|\mathbf{y}_r ({\hat l}) - \boldsymbol{a}_r^H(\theta_{\hat l}^r)\mathbf{H}{\boldsymbol{a}_t(\theta^t)}\sqrt{P_s}s_{\hat m} \right|^2=\\
\left|\sqrt{P_s}h_{l}\left(s_m-s_{\hat m}\right)+\boldsymbol{a}_r^H \left(\theta_{l}^r\right)\mathbf{n}\right|^2.
\end{cases}
\end{equation}
Substituting Eq. (\ref{rorth}) into Eq. (\ref{pr1}), the CPEP can be evaluated as
\begin{equation}\label{sdfg}
\begin{aligned}
P_e
=&\Pr\left(-|\sqrt{P_s}h_{l}\left(s_m-s_{\hat m}\right)|^2\right.\\&\left.
>\left|2\boldsymbol{a}_r^H \left(\theta_{l}^r\right)\mathbf{n}\sqrt{P_s}h_{l}\left(s_m-s_{\hat m}\right)\right|^2\right)\\
=&\Pr\left(-|\sqrt{P_s}h_{l}\left(s_m-s_{\hat m}\right)|^2\right.\\&\left.
-2\Re\left\{n_r\sqrt{P_s}h_{l}\left(s_m-s_{\hat m}\right)\right\}>0\right)\\
=&\Pr\left(G>0\right),
\end{aligned}
\end{equation}
where $G = -|\sqrt{P_s}h_{l}\left(s_m-s_{\hat m}\right)|^2
-2\Re\left\{n_r\sqrt{P_s}h_{l}\left(s_m-s_{\hat m}\right)\right\}$.
Herein, $G$ follows $\mathcal{N}(\mu_G,\sigma_G^2)$ with
$\mu_G=-|\sqrt{P_s}h_{l}\left(s_m-s_{\hat m}\right)|^2$ and
$\sigma_G^2 = |2N_0\sqrt{P_s}h_{l}\left(s_m-s_{\hat m}\right)|^2$.

According to $\Pr(G>0) = Q\left(-{\mu_G}/{\sigma_G}\right)$, the Eq. (\ref{sdfg}) can be further expressed as
{
\begin{equation}\label{qqq}
\begin{aligned}
&P_e=Q\left(\sqrt{\frac{\rho |h_l|^2|s_m - s_{\hat m}|^2}{2}}\right),
\end{aligned}
\end{equation}
}
where $\rho = {P_s}/{N_0}$ denotes average SNR.
\subsubsection{$\hat{l} \neq l$}
In this case, the transmission direction is detected incorrectly, and the erroneous bits originate from both the beam and signal components.
Consequently, the CPEP can be written as
\begin{equation}\label{lnelpr}
\begin{aligned}
P_e=&\Pr\left(\left[{l,m}\right] \rightarrow[\hat{l}, \hat{m}] \right)\\
=&\Pr\left(|\mathbf{y}_r (l) - \boldsymbol{a}_r^H(\theta_{ l}^r)\mathbf{H}\boldsymbol{a}_t(\theta^t)\sqrt{P_s}s_m |^2\right.\\&\left.
>|\mathbf{y}_r ({\hat l}) - \boldsymbol{a}_r^H(\theta_{\hat l}^r)\mathbf{H}\boldsymbol{a}_t(\theta^t)\sqrt{P_s}s_{\hat m} |^2\right).
\end{aligned}
\end{equation}
Depending on Eqs. (\ref{ortho}) and (\ref{lnelpr}), we can obtain
\begin{equation}\label{rorthh}
\begin{cases}
\mathbf{y}_{{r}}(\hat l)
= \mathbf{R}^H  \mathbf{y}(\hat l,\hat m)
=
\boldsymbol{a}_r^H(\theta_{\hat l}^r)\mathbf{n},\\
|\mathbf{y}_r ({l}) - \boldsymbol{a}_r^H(\theta_{ l}^r)\mathbf{H}{\boldsymbol{a}_t(\theta^t)}\sqrt{P_s}s_m |^2
=|\boldsymbol{a}_r^H(\theta_l^r)\mathbf{n}|^2,\\
|\mathbf{y}_r ({\hat l}) - \boldsymbol{a}_r^H(\theta_{\hat l}^r)\mathbf{H}{\boldsymbol{a}_t(\theta^t)}\sqrt{P_s}s_{\hat m}|^2=\\
|\boldsymbol{a}_r^H(\theta_{\hat{l}}^r) \mathbf{n}-\sqrt{P_s}h_{\hat l}s_{\hat{m}}|^2.
\end{cases}
\end{equation}
Combining  Eqs. (\ref{lnelpr}) and (\ref{rorthh}), we can rewritten Eq. (\ref{lnelpr}) as
\begin{equation}\label{eq22}
\begin{aligned}
P_e
=&\Pr\left(|\boldsymbol{a}_r^H(\theta_l^r)\mathbf{n}|^2
>|\boldsymbol{a}_r^H(\theta_{\hat{l}}^r) \mathbf{n}-\sqrt{P_s}h_{\hat l}s_{\hat{m}}|^2\right).
\end{aligned}
\end{equation}
For brevity, let us define $\varrho_1$ and $\varrho_2$ in Eq. (\ref{eq22}) as
$\varrho_1 = |\boldsymbol{a}_r^H(\theta_l^r)\mathbf{n}|^2$ and
$\varrho_2 =|\boldsymbol{a}_r^H(\theta_{\hat{l}}^r) \mathbf{n}-\sqrt{P_s}h_{\hat l}s_{\hat{m}}|^2$,
where $\boldsymbol{a}_r^H(\theta_l^r)\mathbf{n} \sim \mathcal{CN}(0,N_0)$ and $\left(\boldsymbol{a}_r^H(\theta_{\hat{l}}^r) \mathbf{n}-\sqrt{P_s}h_{\hat l}s_{\hat{m}} \right) \sim \mathcal{CN}(-\sqrt{P_s}h_{\hat l}s_{\hat{m}},N_0)$.
{Note that ${\varrho_1}$ denotes the central Chi-squared random variable with two degrees of freedom (DoFs), and ${\varrho_2}$ represents the non-central Chi-squared random variable with two DoFs}.
To facilitate the solution, we make ${x_1 = \frac{\varrho_1}{N_0/2}}$ and
$x_2 = \frac{\varrho_2}{N_0/2}$.
As such, the Eq. (\ref{eq22}) can be calculated as
\begin{equation}\label{np1}
\begin{aligned}
P_e=&\int_0^\infty\int_{x_2}^\infty f_1(x_1) f_2(x_2)dx_1 d x_2,
\end{aligned}
\end{equation}
where the central Chi-square distribution can be described as \cite{kopka}
\begin{equation}\label{np2}
f_1(x_1)=\frac{1}{2}\exp\left(-\frac{x_1}{2}\right).
\end{equation}
Substituting Eq. (\ref{np2}) into Eq. (\ref{np1}), the $P_e$ can be evaluated as
\begin{equation}\label{np3}
\begin{aligned}
P_e
=&\frac{1}{2}\int_0^\infty f_2(x_2)\left(\int_{x_2}^\infty \exp\left(-\frac{x_1}{2}\right) dx_1\right) d x_2.\\
\end{aligned}
\end{equation}
Further, the (\ref{np3}) can be expressed as
\begin{equation}\label{np4}
\begin{aligned}
P_e
=&\int_0^\infty f_2(x_2) \exp\left(-\frac{x_2}{2}\right) d x_2.
\end{aligned}
\end{equation}
Besides, the non-central Chi-square distribution can be calculated as \cite{kopka}
\begin{equation}\label{f2x2}
f_2(x_2)=\frac{1}{2}{\exp\left(-\frac{x_2+\tau}{2}\right)}I_0(\sqrt{kx}),
\end{equation}
where $\tau = 2\rho|h_{\hat{l}}|^2|s_{\hat{m}}|^2$ and $I_0(\cdot)$ can be given as
\begin{equation} \label{I01}
\begin{aligned}
I_0(\sqrt{kx})&=\sum_{k=0}^\infty\frac{\left(\frac{kx}{4}\right)^k}{k!\Gamma(k+1)}\overset{(a)}{=}\sum_{k=0}^\infty\frac{\left(\frac{kx}{4}\right)^k}{k!k!},
\end{aligned}
\end{equation}
where $(a)$ stands for
\begin{equation}\label{gamma}
\Gamma(k+1)=\int_0^\infty t^{k}e^{-x}dx=k!.
\end{equation}
Combining Eqs. (\ref{f2x2}) and (\ref{I01}), the Eq. (\ref{f2x2}) can be evaluated as
\begin{equation}\label{npep3}
f_2(x_2)=\frac{ x_2^k}{2}{\exp\left(-\frac{x_2}{2}\right)}{\exp\left(-\frac{\tau}{2}\right)}\sum\limits_{k=0}^\infty
{\left(\frac{\tau}{4}\right)^k}\frac{1}{k!k!}.
\end{equation}
Then, we substitute Eq. (\ref{npep3}) into Eq. (\ref{np4}), then the
CPEP in Eq. (\ref{np4}) can be obtained as
\begin{equation}\label{npep4}
\begin{aligned}
P_e
=&\frac{1}{2}{\exp\left(-\frac{\tau}{2}\right)}\sum\limits_{k=0}^\infty
{\left(\frac{\tau}{4}\right)^k}\frac{1}{k!k!}\int_0^\infty x_2^k \exp\left(-{x_2}\right) d x_2.\\
\end{aligned}
\end{equation}
Recall that the Gamma function is denoted in Eq. (\ref{gamma}).
At this time, the CPEP can be reformulated as
\begin{equation}\label{npep5}
\begin{aligned}
P_e
=&\frac{1}{2}{\exp\left(-\frac{\tau}{2}\right)}\sum\limits_{k=0}^\infty
{\left(\frac{\tau}{4}\right)^k}\frac{1}{k!}.\\
\end{aligned}
\end{equation}
By utilizing $\exp(x)=\sum_{k=0}^\infty\frac{x^k}{k!}$, the CPEP can be obtained as
\begin{equation}\label{cpep2}
P_e=\frac{1}{2}{\exp\left(-\frac{\rho|h_{\hat{l}}|^2|s_{\hat{m}}|^2}{2}\right)}.
\end{equation}

\subsection{PDF-Based Solution for the UPEP}
It can be observed that the Eq. (\ref{qqq}) contains Q-function, which do not satisfy the form of closed-form expression.
To alleviate this issue, we further derive the closed-form UPEP expressions via the PDF method.
Let us define $x = |h_l|^2$, the UPEP can be derived as follows:
\subsubsection{$\hat{l} = l$}
Combining Eq. (\ref{pdf11}) and $Q(x) = \frac{1}{\pi}\int_0^{\frac{\pi}{2}}\exp\left(-\frac{x^2}{2\sin^2\theta}\right)d\theta$ into Eq. (\ref{qqq}), the UPEP can be calculated as
\begin{equation}\label{ana11}
\bar{P}_e
=  \int_0^\infty f(x)Q\left(\sqrt{\frac{\rho |s_m - s_{\hat m}|^2x}{2}}\right)dx.
\end{equation}
Herein, the $f(x)$ can be derived as follows:
\begin{lemma}
The PDF of $x$ in the descending order can be given by
\begin{equation}\label{bino2}
f(x)
=\frac{L!}{(l-1)!}\sum_{k=0}^{L-l}
\frac{(-1)^k}{k!(L-l-k)!}{\exp\left(-kx-lx\right)}.
\end{equation}
\end{lemma}
\emph{Proof:} Please refer Appendix B for the detail proof.
$\hfill\blacksquare$

Insert Eq. (\ref{bino2}) into Eq. (\ref{ana11}), the UPEP can be written as
\begin{equation}\label{xxww}
\begin{aligned}
\bar{P}_e
=&\frac{L!}{\pi(l-1)!}\sum_{k=0}^{L-l}\frac{(-1)^k}{k!(L-l-k)!}
\int_0^\infty\int_0^{\frac{\pi}{2}} {\exp\left(-kx-lx\right)}\\&\times\exp\left(-\frac{\rho |s_m - s_{\hat m}|^2x}{4\sin^2\theta}\right)d\theta dx.
\end{aligned}
\end{equation}
Exchange the order of integration of Eq. (\ref{xxww}), the UPEP can be further expressed as
\begin{equation}\label{xwefef}
\begin{aligned}
\bar{P}_e
=&\frac{L!}{\pi(l-1)!}\sum_{k=0}^{L-l}\frac{(-1)^k}{k!(L-l-k)!}
\int_0^{\frac{\pi}{2}}\int_0^\infty \\&\times{\exp\left(-\frac{4(k+l)\sin^2\theta+\rho |s_m - s_{\hat m}|^2}{4\sin^2\theta}x\right)}dx d\theta.
\end{aligned}
\end{equation}
Addressing the inner integral of Eq. (\ref{xwefef}), we obtain
\begin{equation}\label{bino4}
\begin{aligned}
\bar{P}_e
=&\frac{L!}{\pi(l-1)!}\sum_{k=0}^{L-l}\frac{(-1)^k}{k!(L-l-k)!(k+l)}\\&\times
\int_0^{\frac{\pi}{2}}{\frac{\sin^2\theta}{\sin^2\theta+\frac{\rho |s_m - s_{\hat m}|^2}{4(k+l)}}} d\theta.
\end{aligned}
\end{equation}
After some mathematical operations, the Eq. (\ref{bino4}) can be reformulated as
\begin{equation}\label{bino5}
\begin{aligned}
\bar{P}_e
=&\frac{L!}{2(l-1)!}\sum_{k=0}^{L-l}\frac{(-1)^k}{k!(L-l-k)!(k+l)}\\&\times\left(1-\sqrt{\frac{\rho |s_m - s_{\hat m}|^2}{\rho |s_m - s_{\hat m}|^2+4(k+l)}}\right).
\end{aligned}
\end{equation}

\subsubsection{$\hat{l} \neq l$}
In this case, combing Eqs. (\ref{bino2}) and (\ref{ana11}), we can get
\begin{equation}\label{eq44}
\begin{aligned}
\bar{P}_e
=  &\frac{1}{2}\int_0^\infty f(x)\exp\left(-\frac{\rho |s_{\hat m}|^2}{2}x\right)dx\\
= & \frac{1}{2}\int_0^\infty \frac{L!}{(\hat{l}-1)!}\sum_{k=0}^{L-\hat{l}}
\frac{(-1)^k}{k!(L-\hat{l}-k)!}{\exp\left(-kx-\hat{l}x\right)}\\&\times\exp\left(-\frac{\rho |s_{\hat m}|^2}{2}x\right)dx\\
= & \frac{L!}{(\hat{l}-1)!}\sum_{k=0}^{L-\hat{l}}
\frac{(-1)^k}{k!(L-\hat{l}-k)!(2k+2\hat{l}+\rho|s_{\hat m}|^2)}.
\end{aligned}
\end{equation}
\subsection{MGF-based Solution for the UPEP }
In this subsection, we aim to derive the UPEP closed-form expression for the proposed scheme with an alternative approach.
In the {\bf Lemma 3}, we provide the MGF expression to evaluate the UPEP of the proposed RIS-SSM scheme.

\begin{lemma}
Since variable $|h_l|, l\in\{1,2,\cdots,L\}$, sorts in the descending order, the corresponding MGF of $x=|h_l|^2$ can be given by
\begin{equation}\label{MGFw1}
{\rm MGF}_x(s)
=\prod \limits_{\xi=l}^L\frac{\xi}{\xi-s}.
\end{equation}
\end{lemma}
\emph{Proof:} Please refer Appendix C for the detail proof.$\hfill\blacksquare$

Based on MGF expression, the UPEP can be derived as follows:
\subsubsection{$\hat{l} = l$}
Substitute Eq. (\ref{MGFw1}) into Eq. (\ref{ana11}), we evaluate the UPEP as
\begin{equation}\label{mgfww}
\begin{aligned}
\bar{P}_e
= & \int_0^\infty f(x)Q\left(\sqrt{\frac{\rho |s_m - s_{\hat m}|^2x}{2}}\right)dx \\
=&\frac{1}{\pi}\int_0^{\frac{\pi}{2}}\int_0^\infty f(x)\exp\left(-\frac{\rho |s_m - s_{\hat m}|^2x}{4\sin^2\theta}\right)dx d\theta\\
=&\frac{1}{\pi}\int_0^{\frac{\pi}{2}}\prod \limits_{\xi=l}^L\frac{{\sin^2\theta}}{{\sin^2\theta}+{\frac{\rho |s_m - s_{\hat m}|^2}{4\xi}}}d\theta.
\end{aligned}
\end{equation}
Let us set $\nu = \xi -l+1$, the Eq. (\ref{mgfww}) can be reformulated as
\begin{equation}\label{mgfww2}
\bar{P}_e
=\frac{1}{\pi}\int_0^{\frac{\pi}{2}}\prod \limits_{\varsigma=1}^{L-l+1}\frac{{\sin^2\theta}}{{\sin^2\theta}+{\frac{\rho |s_m - s_{\hat m}|^2}{4(\varsigma+l-1)}}}d\theta.
\end{equation}
After some calculations, we can get
\begin{equation}
\begin{aligned}
\bar{P}_e
&=\frac{1}{2}\sum_{\varsigma=1}^{\dot{L}}\left(1-\sqrt{\frac{c_{\varsigma}}{1+c_{\varsigma}}}\right)
\prod\limits_{\nu=1,n\neq \varsigma}^{\dot L}\left(\frac{c_{\varsigma}}{c_{\varsigma}-c_{\nu}}\right),
\end{aligned}
\end{equation}
where
$\dot{L}=L-l+1$, $c_{\varsigma}={\frac{\rho |s_m - s_{\hat m}|^2}{4(\varsigma+l-1)}}$, and $c_{\nu}={\frac{\rho |s_m - s_{\hat m}|^2}{4(\nu+l-1)}}$.

\subsubsection{$\hat{l} \neq l$}
Based on Eq. (\ref{eq44}), we have
\begin{equation}\label{upepee}
\begin{aligned}
\bar{P}_e
&=  \frac{1}{2}\int_0^\infty f(x)\exp\left(-\frac{\rho |s_{\hat m}|^2}{2}x \right)dx\\&
=\frac{1}{2}\prod \limits_{\xi={\hat l}}^L\frac{2\xi}{2\xi+\rho|s_{\hat m}|^2}.
\end{aligned}
\end{equation}

Note that both PDF and MGF methods apply the equivalent transformed form of the Q-function.
However, when the form of the integral function is very complex, we usually apply the approximate form of the Q-function to obtain the closed-form UPEP.
In this case, the UPEP derived via the Q-function can be given in the following theorem.

\begin{theorem}
It is a common approach to derive a closed-form expression for UPEP based on  $Q(x) \approx \frac{1}{12}\exp\left(-\frac{x^2}{2}\right)+\frac{1}{4}\exp\left(-\frac{2x^2}{3}\right)$ {\rm\cite{ma2020large,li2021space}} as follows:
\begin{equation}\label{mgfbasedd}
\bar{P}_e=
\begin{cases}
\frac{1}{12}\prod \limits_{\xi=l}^L\left(\frac{48\xi^2+13\rho |s_m - s_{\hat m}|^2\xi}{12\xi^2+7\rho |s_m - s_{\hat m}|^2\xi+({\rho |s_m - s_{\hat m}|^2})^2}\right), &\hat l = l,\\
\frac{1}{2}\prod \limits_{\xi={\hat l}}^L\frac{2\xi}{2\xi+\rho|s_{\hat m}|^2}, &\hat l \neq l.
\end{cases}
\end{equation}

\end{theorem}
{\emph{Proof:}} Please refer Appendix D for the detail proof. $\hfill\blacksquare$
\subsection{Asymptotic UPEP}
This subsection investigates the asymptotic UPEP in the high SNR region, where the mathematical derivation of asymptotic UPEP is as follows.
\subsubsection{$\hat{l} = l$}
Recalling Eq. (\ref{mgfww}), the asymptotic expression of the UPEP in the high SNR region can be evaluated as
\begin{equation}\label{walli1}
\begin{aligned}
\bar{P}_{\rm asy}
=&\frac{1}{\pi}\int_0^{\frac{\pi}{2}}\prod \limits_{\xi=l}^L\frac{{\sin^2\theta}}{{\sin^2\theta}+{\frac{\rho |s_m - s_{\hat m}|^2}{4\xi}}}d\theta\\
\approx &\frac{1}{\pi}\int_0^{\frac{\pi}{2}}\prod \limits_{\xi=l}^L\frac{{\sin^2\theta}}{{\frac{\rho |s_m - s_{\hat m}|^2}{4\xi}}}d\theta\\
=&\frac{1}{\pi}\prod \limits_{\xi=l}^L\left(\frac{4\xi}{{\rho |s_m - s_{\hat m}|^2}}\right)\int_0^{\frac{\pi}{2}}{{(\sin\theta)^{2\dot{L}}}}d\theta.
\end{aligned}
\end{equation}
Herein, we resort to the Wallis formula \cite{jef2007book}, the asymptotic expression of Eq. (\ref{walli1}) can be further expressed as
\begin{equation}\label{eq53}
\begin{aligned}
\bar{P}_{\rm asy}
&=\frac{1}{\pi}\prod \limits_{\xi=l}^L\left(\frac{4\xi}{{\rho |s_m - s_{\hat m}|^2}}\right)\frac{(2\dot L-1)!!}{(2\dot L)!!}\frac{\pi}{2}\\&
=\frac{(2\dot L-1)!!}{2(2\dot L)!!}\frac{\Gamma(L+1)}{\Gamma(l)}\left(\frac{4}{{\rho|s_m - s_{\hat m}|^2}}\right)^{\dot L}.
\end{aligned}
\end{equation}
\subsubsection{$\hat{l} \neq l$}
Based on Eq. (\ref{upepee}), following the similar process in Eq. (\ref{eq53}), the asymptotic UPEP can be calculated as
\begin{equation}\label{lneqlll}
\begin{aligned}
\bar{P}_{\rm asy}
=&\frac{1}{2}\prod \limits_{\xi={\hat l}}^L\frac{1}{1+\frac{\rho| s_{\hat m}|^2}{2\xi}}
\approx \frac{2^{\ddot L}}{2}\prod \limits_{\xi={\hat l}}^L\frac{\xi}{{\rho| s_{\hat m}|^2}}\\
=&\frac{\Gamma(L+1)}{2\Gamma({\hat l})}\left(\frac{2}{{\rho| s_{\hat m}|^2}}\right)^{\ddot L},
\end{aligned}
\end{equation}
where $\ddot L = L-\hat{l}+1$.

\subsection{ABEP Expression}
Based on the UPEP, the union upper bound of ABEP of the proposed RIS-SSM scheme can be written as
\begin{equation}\label{abep}
\begin{aligned}
{\rm ABEP} \leq& \frac{1}{ML_s\log_2({ML_s})}\sum\limits_{l=1}^L\sum\limits_{m=1}^M\sum\limits_{\hat l=1}^L\sum\limits_{\hat m=1}^M\bar{P}_e\\&\times N(\left[{l}, {m}\right] \rightarrow[\hat{l}, \hat{m}] ),
\end{aligned}
\end{equation}
where $N(\left[{l}, {m}\right] \rightarrow[\hat{l}, \hat{m}])$ stands for the Hamming distance between the binary expressions of the symbols $\hat l$ to $l$ and $\hat m$ to $m$.

\subsection{Diversity Gain }
In view of the MGF-based approach to evaluating
ABEP discussed in the previous subsection, we investigate the diversity gain of the proposed RIS-SSM scheme in this subsection.
\subsubsection{$\hat l = l$}
Substituting Eq. (\ref{eq53}) into the operation, the diversity gain $\mathcal{D}$ can be formulated as \cite{zheng2003diver}
\begin{equation}\label{revel}
\begin{aligned}
\mathcal{D}=&\lim\limits_{\rho \to \infty}-\frac{\log_2({\rm ABEP})}{\log_2\rho}\\
=&\lim\limits_{\rho \to \infty}-\frac{\log_2\left({\frac{1}{2}\frac{(2\dot L-1)!!}{(2\dot L)!!}\frac{\Gamma(L+1)}{\Gamma(l)}\left(\frac{4}{{\rho|s_m - s_{\hat m}|^2}}\right)^{\dot L}}\right)}{\log_2\rho}.
\end{aligned}
\end{equation}
Ignoring the part not related to  $\rho$, we can obtain
\begin{equation}\label{fig59}
\begin{aligned}
\mathcal{D}
=&\lim\limits_{\rho \to \infty}-\frac{{\dot L}\log_2{\left(\frac{4}{{\rho|s_m - s_{\hat m}|^2}}\right)}}{\log_2\rho}.
\end{aligned}
\end{equation}
After some manipulations, Eq. (\ref{fig59}) can be further simplified to
\begin{equation}\label{diver1}
\begin{aligned}
\mathcal{D}
=&\lim\limits_{\rho \to \infty}\frac{{\dot L}\log_2{{{\rho}}}+2{\dot L}\log_2{\left({{|s_m - s_{\hat m}|}}\right)}-2\dot L }{\log_2\rho}
= \dot L.
\end{aligned}
\end{equation}

\subsubsection{$\hat l \neq l$}
In this case, the diversity gain can be given as
\begin{equation}
\begin{aligned}
\mathcal{D}=&\lim\limits_{\rho \to \infty}-\frac{\log_2({\rm ABEP})}{\log_2\rho}\\
=&\lim\limits_{\rho \to \infty}-\frac{\log_2\left(\frac{\Gamma(L+1)}{\Gamma({\hat l})}\left(\frac{2}{{\rho| s_{\hat m}|^2}}\right)^{\ddot L}\right)}{\log_2\rho}.
\end{aligned}
\end{equation}
Further, $\mathcal{D}$ can be expressed as
\begin{equation}\label{diver2}
\begin{aligned}
\mathcal{D}
=&\lim\limits_{\rho \to \infty}-\frac{\log_2\left(\frac{\Gamma(L+1)}{\Gamma({\hat l})}\right)+{\ddot L}\log_2\left(\frac{2}{{\rho| s_{\hat m}|^2}}\right)}{\log_2\rho}\\
=&\lim\limits_{\rho \to \infty}-\frac{{\ddot L}-{\ddot L}\log_2\rho-{\ddot L}\log_2\left({{| s_{\hat m}|^2}}\right)}{\log_2\rho}
= \ddot L.
\end{aligned}
\end{equation}
Due to $l, \hat{l} \in \{1,2,\cdots,L_s\}$,  we can combine Eqs. (\ref{diver1}) and (\ref{diver2}) to obtain diversity gain as
\begin{equation}\label{divsigain}
\mathcal{D} = L-L_s+1.
\end{equation}
From (\ref{divsigain}), we can observe that the diversity gain of the proposed RIS-SSM is only related to $L$ and $L_s$.

\subsection{Ergodic Capacity}
In the proposed RIS-SSM scheme, discrete constellation points are used to transmit bits of information.
In light of this, we evaluate the discrete-input continuous-output memoryless channel (DCMC) capacity, which can be written as
\begin{equation}
C = \mathbb{E}_{h_l}\left[\max\limits_{h_l,s_m}
\mathcal{I}(s_m,h_l;\mathbf{y}_r(l) )\right],
\end{equation}
where $\mathcal{I}(h_l,s_m;\mathbf{y}_r(l))$ denotes the mutual information between the input signal $\{h_l,s_m\}$ and received signal $\mathbf{y}_r(l)$.

\begin{proposition}
The lower-bound of the ergodic capacity for the proposed scheme can be given as
\begin{equation}\label{theore1}
\begin{aligned}
C &= 2\log_2(L_sM)-\log_2\left[L_sM\right.\\&\left.+\sum_{m=1}^M\sum_{l=1}^{L_s}\sum_{\hat{m}\neq m = 1}^M\sum_{\hat{l} \neq l = 1}^{L_s}\frac{\Gamma(L+1)}{\Gamma({\hat l})}\left(\frac{2}{{\rho| s_{\hat m}|^2}}\right)^{\ddot L}
\right].
\end{aligned}
\end{equation}
\end{proposition}
{\emph {Proof:}}  Please refer Appendix E for the detail proof.$\hfill\blacksquare$

\subsection{System Throughput}
The system throughput represents a measure of the information from Tx to Rx of the system. To be precise, system throughput stands for the maximum data rate at which Rx is successfully reached.
Thus, the system throughput of RIS-SSM scheme can be given as

\begin{equation}
ST = {\rm(1-ABEP)} \log_2(L_sM).
\end{equation}

\section{Numerical and Analytical Results}
In this section, simulation results and analytical derivations of the RIS-SSM scheme are provided.
{For the simulation, the channel implementation is randomly generated $1 \times 10^6$ times and then obtained by averaging the operations.
We have not consider the pilot power overhead in the following simulations.
If the pilot power is considered practically, all SNR-related figures will be shifted as a whole in the direction of the larger SNR, and the moved distance is related to the pilot power over the total power.
}

\subsection{ Validation of Analytical Results}

\begin{figure}[t]
\centering
\subfigure[$L = 4, L_s = 2$, and $M = 2$.]
{
\begin{minipage}[b]{0.22\textwidth}
\includegraphics[width=4.4cm]{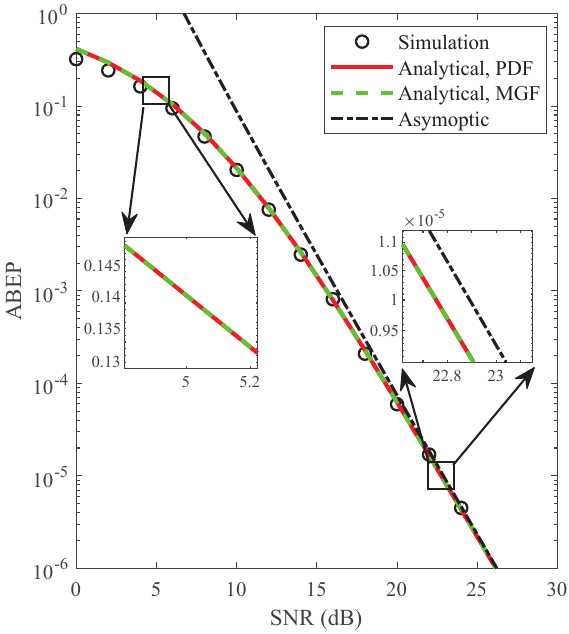}
\end{minipage}
}
\subfigure[$L = 18, M = 4$, and SNR = 24 dB.]
{
\begin{minipage}[b]{0.22\textwidth}
\includegraphics[width=4.4cm]{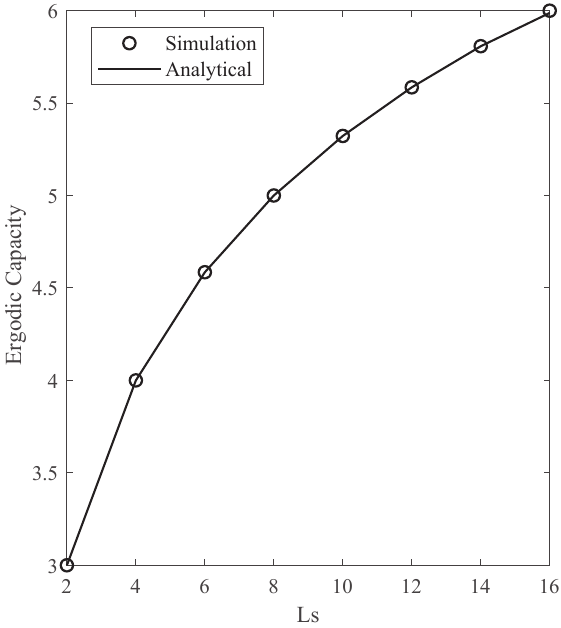}
\end{minipage}
}
\caption{\small{Verification of the analytical derivations for the proposed RIS-SSM scheme.}}
\label{figverf}
\end{figure}
\subsubsection{ABEP}
In Fig. {\ref{figverf}}(a), we utilize Monte Carlo simulation to verify the analytical ABEP expressions of the proposed RIS-SSM scheme. Analytical results are composed of union upper bound and asymptotic expression of ABEP, where the union upper bound of ABEP is derived via the PDF-based method and MGF-based method. In this figure, we set the transmission rate as 2 bit/s/Hz, where the total number of scatterers is 4.
For the union upper bound of ABEP, our analysis results are in perfect agreement with the simulation values in the high SNR region. In contrast, the analysis results show slight differences especially in the low SNR region.
This phenomenon fully demonstrates the validity of the results of our analysis.
In particular, we can observe from the two magnifications that the two approaches derive precisely the same results.
For asymptotic ABEP, as the SNR increases, the trend remains the same as for the union upper bound. In the enlarged figure on the right of Fig. {\ref{figverf}}(a), it is clear that the value is slightly larger than the union upper bound of ABEP, which is completely consistent with our expectation.

\subsubsection{Ergodic Capacity}
In Fig. {\ref{figverf}}(b), we present the ergodic capacity versus the various $L_s$ values, where Monte Carlo simulations and analytical ergodic capacity results are provided by Eqs. (\ref{egrodi1}) and (\ref{theore1}), respectively.
In Fig. {\ref{figverf}}(b), we assume the number of scatterers is 18 and the modulation method employed in the symbol domain is QPSK.
It can be observed that the simulation results and the analytical curves match perfectly, which verifies the correctness of the derivations.
Noteworthy, as $L_s$ increases, the increment of ergodic capacity gradually decreases.
Recall that amplitude gains of the scatterers involved in the modulation sort in descending order.
As the modulation path increases, the corresponding scatterer gain gradually decreases.

\begin{figure}[t]
\centering
\subfigure[$L = 6, L_s = 4,$ and $M = 4$.]
{
\begin{minipage}[b]{0.22\textwidth}
\includegraphics[width=4.4cm]{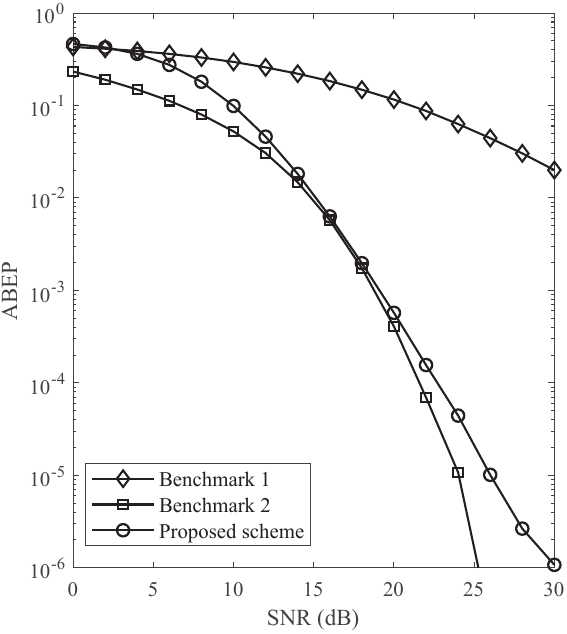}
\end{minipage}
}
\subfigure[$L = 12, L_s = 4,$ and $M = 4$.]
{
\begin{minipage}[b]{0.22\textwidth}
\includegraphics[width=4.4cm]{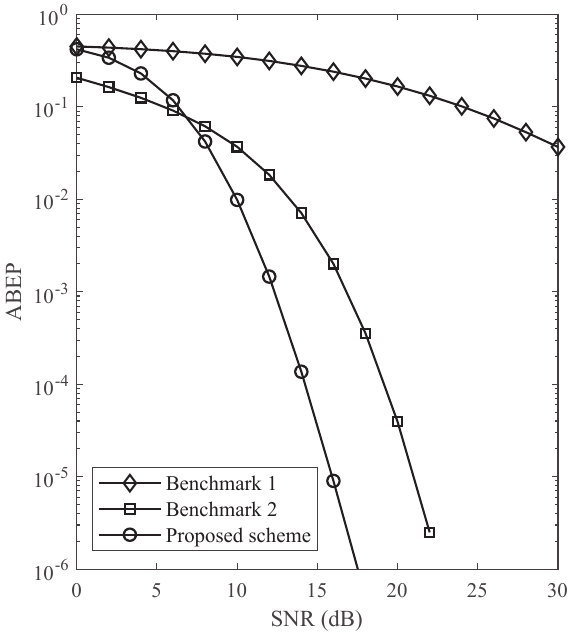}
\end{minipage}
}
\caption{\small{Comparison with benchmark 1 and benchmark 2. }}
\label{benchmark12}
\end{figure}

\subsection{Comparison with Benchmarks}

In Fig. \ref{benchmark12}, we compare the proposed scheme with two benchmarks: 1) Benchmark 1: the reflected beam steering to the scatterer with the lowest gain  $h_L$ in the RIS-Rx channel. 2) Benchmark 2: the reflected beam steering to the scatterer with the highest gain  $h_1$ in the RIS-Rx channel. For a fair comparison, we set the data rate as 4 bit/s/Hz, where the RIS-SSM scheme utilizes QPSK in the symbol domain while the two benchmarks use 16QAM.
As shown in Fig. \ref{benchmark12}, the RIS-SSM scheme achieves better ABEP performance when the total number of scatterers becomes larger. Moreover, since benchmark 1 and benchmark 2 respectively denote the instantaneous minimum and maximum SNR beams. That is, there are no error bits generated in the spatial domain, all the information bits are produced in the symbol domain.
From Fig. \ref{benchmark12}(a), we can see that the performance of the RIS-SSM system is between benchmark 1 and benchmark 2.
Conversely, it is evident in Fig. \ref{benchmark12}(b) that the ABEP performance of the proposed scheme is significantly better than the two benchmarks.
Because the scatterer amplitude gains are in descending order, with the total number of scatterers increases, the sample table scatterer gain of Fig. \ref{benchmark12}(b) becomes more prominent than that of Fig. \ref{benchmark12}(a) in the RIS-SSM system.
It is worth noting that in Fig. \ref{benchmark12}(a), we observed a bend in the proposed scheme and benchmark 2.
Also, the value of benchmark 2 is missing at SNR = 24 dB in Fig. \ref{benchmark12}(b).
Both phenomena are caused by the insufficient number of Monte Carlo simulations. To alleviate this problem, we can increase the number of simulations to $10^7$ or higher.

\begin{figure}[t]
\begin{minipage}[t]{0.49\linewidth}
\centering
\includegraphics[width=4.5 cm]{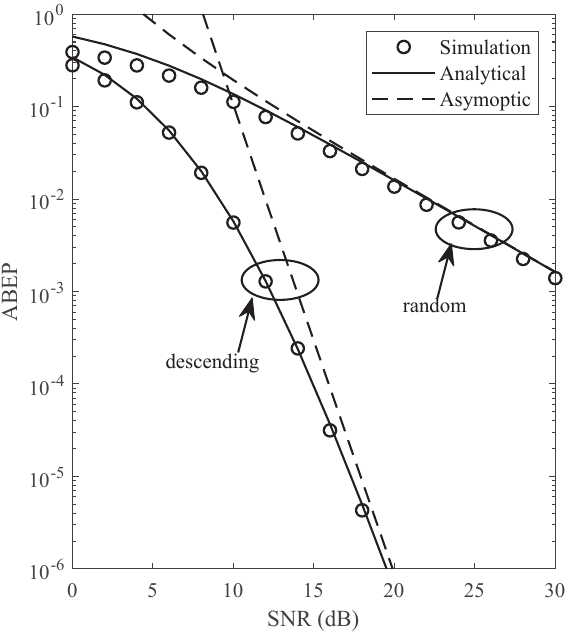}
\caption{\small{ABEP comparison of the RIS-SSM scheme with no sorting scheme.}}
\label{nosort}
\end{minipage}%
\hfill
\begin{minipage}[t]{0.49\linewidth}
\centering
\includegraphics[width=4.5 cm]{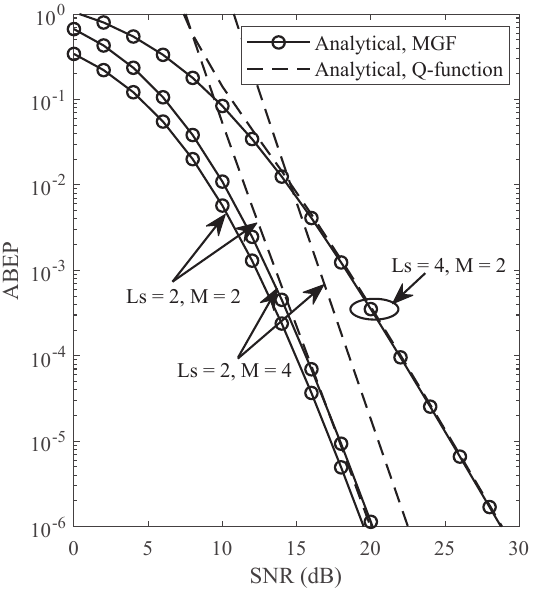}
\caption{\small{ABEP comparison of the RIS-SSM scheme with the Q-function method.}}
\label{qapprox}
\end{minipage}
\end{figure}
\subsection{Comparison of Scatterer Selection Strategies}

In Fig. \ref{nosort}, we compare the scatterer gains in the RIS-SSM scheme in a descending order and in a random order.
For a fair comparison, the parameters of the system are set uniformly to $L_s=2, M=2$, and $L$ = 4.
In this figure, the obtained analytical results are compared with the simulation values to confirm the correctness of the derived upper bound and asymptotic curves.
As expected, we find that the performance of selecting $2$ scatterers from a descending order of scatterer gain is significantly better than that of selecting $2$ scatterers from a random order of scatterer gain when the total number of scatterers is $4$.
To be specific, the performance of the system with scatterers arranged in descending order is approximately 10 dB less than that of a random arrangement with ABEP = $10^{-2}$.
This is because, when the scatterer gains are in descending order, the RIS-SSM system can obtain the diversity gain provided by the number of RF chains at the receiving end.
In particular, when $L=2$, the RIS-SSM system obtains the same ABEP under the two strategies of descending magnitude of scatterers and random arrangement. This is because the modulation order of the spatial domain is randomly selected in the range of $L_s = L$ and thus is not limited by the sorting algorithm.

\subsection{Comparison with Approximation Method}
In Fig. \ref{qapprox}, we compare two methods based on MGF and Q-function estimation approaches with respect to analytical ABEP derivation, where the transmission rate is set to 2 bits/s/Hz and 3 bits/s/Hz.
For comparison, 2 bits/s/Hz is the reference. Note that 3 bits/s/Hz consists of QPSK with 2 scatterers and BPSK with 4 scatterers.
According to this figure, there is a noticeable performance degradation between the two approaches, especially when the $M$ increases.
However, as $L_s$ raises, the ABEP difference between Q-function approximation and MGF-based method can be nearly ignorable.
Due to theorem 1, the Q-function estimation approach is adopted by ABEP in the $\hat l = l$ case, while the same expression as MGF is employed in the $\hat l \neq l$ case.
When $M$ increases, the sample range of constellation points for the Q-function estimation approach expands and the ABEP becomes larger. Thus, the gap between two approaches becomes larger. In contrast, when $L_s=4$, the BPSK is employed, the ABEP in the symbol domain remains unchanged.
Furthermore, since the scatterer gains are in descending order, the gain of the Q-function estimation approach increases as $L_s$ improves, the ABEP gap between the two approaches decreases.

\begin{figure}[t]
\centering
\includegraphics[width=4.5cm]{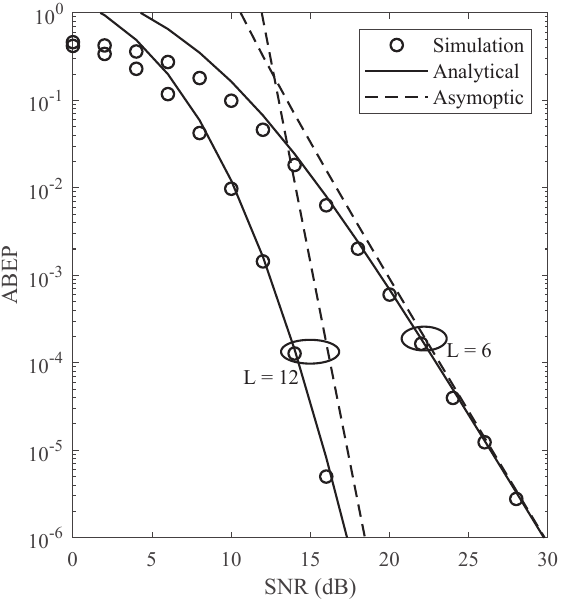}
\caption{\small{Impact of the number of scatterers on the ABEP. }}
\label{figL}
\end{figure}
\subsection{ABEP Performance Evaluation}
\subsubsection{Impact of  $L$ on the ABEP performance}
In Fig. \ref{figL}, we present the ABEP performance of the proposed RIS-SSM scheme for the spectral efficiency of 4 bpcu, where $L_s = 4$ and QPSK modulation is adopted. It is seen from Fig. \ref{figL} that ABEP decreases very rapidly as the total number of scatterers is 12, while ABEP decreases slightly more slowly as the total number of scatterers is 6. Recall that $\mathcal{D}=L-L_s +1$ provided in Eq. (\ref{divsigain}), we can obtain that when $L$ equals 6 and 12 the corresponding diversity gains are 3 and 9, respectively.
The reason is that when the scatterers in the environment are more enriched, the quality of the scatterers selected by the descending order algorithm is higher.

\subsubsection{Impact of  $L_s$ on the ABEP performance}
In Fig. \ref{datarate}(a),  we investigate the impact of the modulation order of the scatterers of the RIS-SSM scheme on the ABEP performance.
As can be seen from Fig. \ref{datarate}(a), as the modulation order $L_s$ decreases, the ABEP performance becomes better.
Due to each transmission, the RIS turns its beam towards fewer scatterers, which means that the modulation order becomes smaller. Euclidean distance between the normalized constellation points becomes more considerable. As a result, the probability of an error in the scatterer's verdict also becomes smaller. In addition, we also find that the slope of ABEP changes with the number of scatterers participating in the modulation because the diversity gain and the modulation order $L_s$ are also related.
Since the number of RF chains at Rx requires to be no less than $L_s$, this condition ensures that Rx can identify the beams from each scatterer to make an effective decoding decision.

\subsubsection{Impact of $M$ on the ABEP performance}
\begin{figure}[t]
\centering
\subfigure[$L = 12, M = 4$.]{
\begin{minipage}[b]{0.22\textwidth}
\includegraphics[width=4.4 cm]{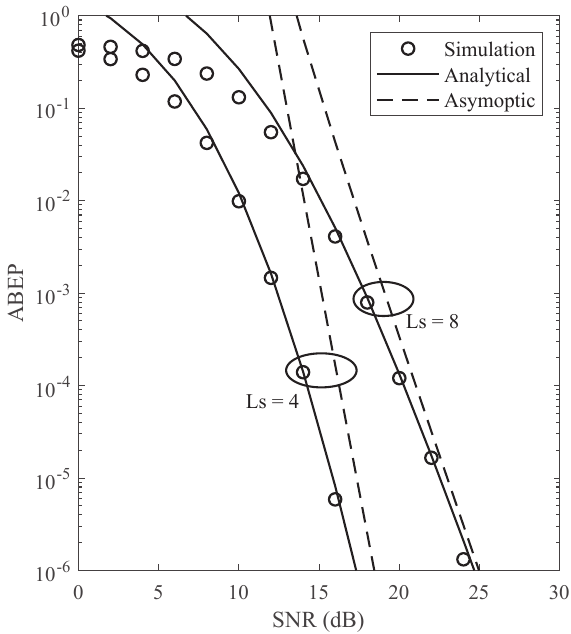}
\end{minipage}
}
\subfigure[$L = 12, L_s = 4$.]{
\begin{minipage}[b]{0.22\textwidth}
\includegraphics[width=4.4 cm]{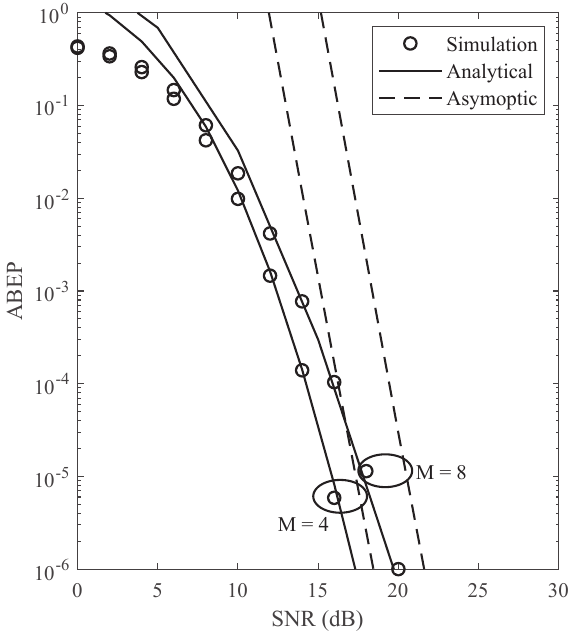}
\end{minipage}
}
\caption{\small{ABEP performance under different data rates. }}
\label{datarate}
\end{figure}

In Fig. \ref{datarate}(b), we compare the ABEP performance of the RIS-SSM scheme for $L=12$ and $L_s=4$ cases, where the $L_s$ with the highest gain are selected in descending order of $L$ scatterer gain to be modulated.
The symbol modulation order $M$ is set to be 4 and 8, respectively.
As shown in Fig. \ref{datarate}(b), the proposed RIS-SSM scheme with $M = 4$ outperforms its counterpart with $M = 8$.
Specifically, at ABEP = $10^{-6}$, system with $M=8$ requires 2 dB more SNR than in the case of $M=4$.
Additionally, it is clear from Fig. \ref{datarate}(b) that the slope of the asymptotic ABEP corresponding to increasing the symbol modulation order $M$ remains unchanged when $L$ and $L_s$ are fixed, which can be explained by our derived diversity gain in Eq. (\ref{divsigain}).
\subsection{Comparison of Ergodic Capacity Under Various $L$}
\begin{figure}[t]
\begin{minipage}[t]{0.49\linewidth}
\centering
\includegraphics[width=4.5cm]{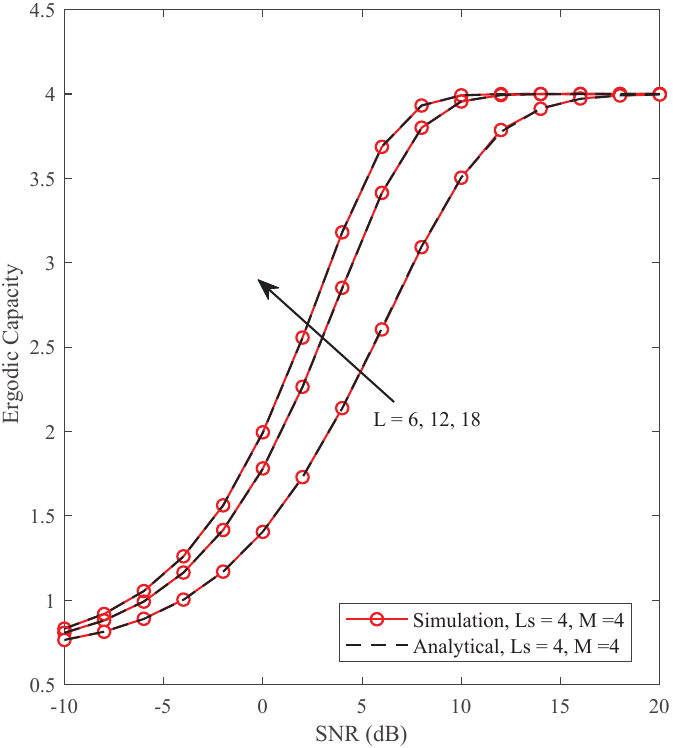}
\caption{\small{Ergodic capacity versus SNR under different $L$.}}
\label{figec}
\end{minipage}%
\hfill
\begin{minipage}[t]{0.49\linewidth}
\centering
\includegraphics[width=4.5cm]{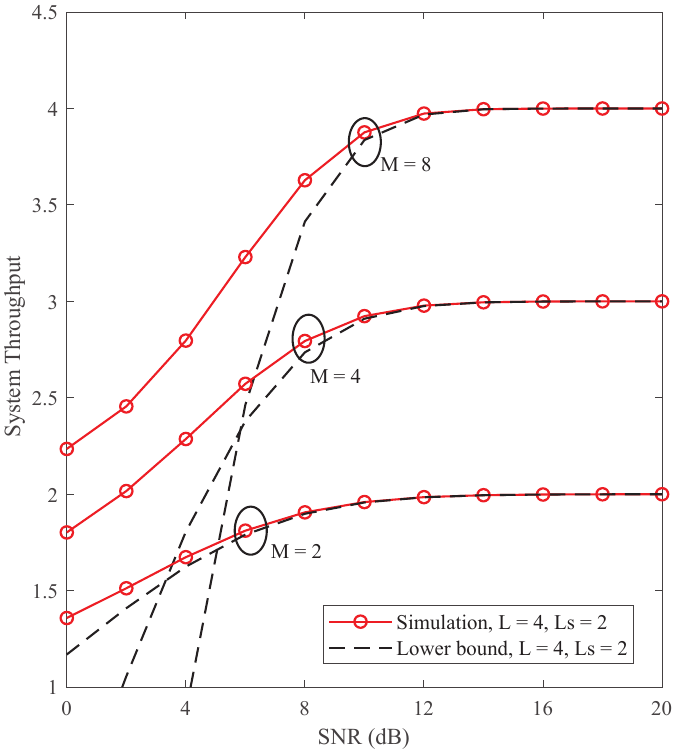}
\caption{\small{System throughput versus SNR under different $M$.}}
\label{figst}
\end{minipage}
\end{figure}

The simulation results and analytical curves on the ergodic capacity of the RIS-SSM scheme are depicted in Fig. \ref{figec}.
In particular, all analytical curves offer a perfect match with simulation results.
It can be seen from the figure that the ergodic capacity increases linearly in lower SNR region and approaches towards the same upper limit in higher SNR region, since the modulation parameters $L_s$ and $M$ remain constant.
Nevertheless, the ergodic capacity of the system still shows distinct styles for different scatterers in the channel.
More specifically,
if the ergodic capacity is 2.5, the systems corresponding to $L=6$ and 12 need approximately 3.6 dB and 1 dB more SNR than the RIS-SSM system corresponding to $L=18$, respectively.
The reason for this phenomenon is that as the total number of scatterers in the channel increases, the selection probability of the higher gain of the candidate scatterers arranged in descending order becomes greater.
\subsection{Comparison of System Throughput Under Various $M$}

Fig. \ref{figst} depicts the lower bound of the system throughput calculated by Eq. (\ref{theore1}) and the exact system throughput obtained by simulations.
Specifically, as the SNR increases, the lower bound of the system throughput converges to the exact system throughput.
It is evident from Fig. \ref{figst} that in the low SNR region, there is a difference between the system throughput of the simulation and the lower limit of the system throughput. In the high SNR region, the two are in high agreement.
Depending on these results, it can be inferred that the lower bound of system throughput at high SNR can be employed to estimate the achievable system throughput of the RIS-SSM system.
Moreover, we observe from Fig. \ref{figst} that the system throughput of the proposed scheme is close to its upper bound of system throughput in the high SNR region (i.e., $\log_2ML_s$).
Based on these observations, it can be inferred that system throughput of the proposed scheme depends on the data rate.

\section{Conclusion and furture works}

In this paper, we investigated the RIS-SSM scheme for the mmWave MIMO communication, where the Tx-RIS and the RIS-Rx channels are adopted LoS and NLoS links, respectively. Based on this architecture, we use the ML detector to derive the CPEP expression from the perspective that the received beam is detected correctly and incorrectly, and then derived closed-form expressions of UPEP via the PDF-based and MFG-based approaches, respectively. In particular, PDF and MGF expressions for descending scatterer gains, asymptotic ABEP expressions, and diversity gains are also provided in the high SNR region. Also, the conventional method based on Q-function approximating to obtain the closed-form expression of UPEP was derived in this paper.
To characterize the capacity of the RIS-SSM system effectively, the exact expression of ergodic capacity and the lower bound expression of system throughput are also given in this work.
The simulation results validated the analytical ABEP, ergodic capacity, and system throughout results, and confirm that the proposed RIS-SSM scheme can obtain better ABEP performance than maximal and minimal beamforming schemes.
In addition, the ABEP results obtained based PDF and MGF methods outperform than the conventional Q-function approximating method.
Furthermore, the richer the scatterer, the better it is to enhance the ergodic capacity of the proposed system.
Overall, this work has provided valuable insights into the performance of the RIS-SSM scheme for mmWave MIMO communication, and the simulation and analytical results presented in this paper can serve as a reference for future research in this field.
{
It is worth mentioning that receivers with interference in RIS or discrete phase shifting in RIS as well as MMSE detection are still interesting and open research issues.
}

\begin{appendices}
\section{}
For the large-scale ULA, the array response vectors of $N_t$-antenna can be expressed as
\begin{equation}\label{ula3}
\begin{aligned}
&\boldsymbol{a}^H(\theta_{l_1})\boldsymbol{a}(\theta_{l_2})\\
= & \frac{1}{\sqrt{N_t}}[1,\exp(-j2\pi\phi_{l_1}),\ldots,\exp(-j2\pi\phi_{l_1}(N_t-1))]\\
&\times \frac{1}{\sqrt{N_t}}[1,\exp(j2\pi\phi_{l_2}),\ldots,\exp(j2\pi\phi_{l_2}(N_t-1))]^T\\
=&\frac{\exp\left(j\pi(\phi_{l_2}-\phi_{l_1})N_t\right)}{N_t\exp\left(j\pi(\phi_{l_2}-\phi_{l_1})\right)}\\&\times\frac{\left(\exp\left(-j\pi(\phi_{l_2}-\phi_{l_1})N_t\right)-\exp\left(j\pi(\phi_{l_2}-\phi_{l_1})N_t\right)\right)}{\left(\exp\left(-j\pi(\phi_{l_2}-\phi_{l_1})\right)-\exp\left(j\pi(\phi_{l_2}-\phi_{l_1})\right)\right)}\\
=&\frac{1}{N_t}\frac{\sin\left(\pi(\phi_{l_2}-\phi_{l_1})N_t\right)}{\sin\left(\pi(\phi_{l_2}-\phi_{l_1})\right)}\exp\left(j\pi(\phi_{l_2}-\phi_{l_1})(N_t-1)\right).
\end{aligned}
\end{equation}
where  $\phi_l={ d\sin\theta_l}/{\lambda}$.
Afterwards, we divide the subsequent proof into $l_1 = l_2$ and $l_1 \neq l_2$ cases.

1) If $l_1 = l_2$, it is easy to have $\boldsymbol{a}^H(\theta_{l_1})\boldsymbol{a}(\theta_{l_2})=1$.

2) If $l_1 \neq l_2$, we define a function as $f(x) = \left|\frac{\sin(N_tx)}{N_t\sin(x)}\right|$. Then, substituting $f(x)$ into $\alpha_{l_1,l_2}$, we have
{
\begin{equation}
\begin{aligned}
f(\pi(\phi_{l_2}-\phi_{l_1}))&=\left|\frac{\sin(N_t\pi(\phi_{l_2}-\phi_{l_1}))}{N_t\sin(\pi(\phi_{l_2}-\phi_{l_1}))}\right|\\
&\approx\left|\frac{\sin(N_t\pi(\phi_{l_2}-\phi_{l_1}))}{N_t\pi(\phi_{l_2}-\phi_{l_1})}\right|\\&
\leq \left|\frac{1}{N_t\pi(\phi_{l_2}-\phi_{l_1})}\right|\\&
=\left|\frac{\lambda}{N_t\pi d(\sin\theta_{l_2}-\sin\theta_{l_1})}\right|.
\end{aligned}
\end{equation}
}
Without loss of generality, let us define
$\Delta = \sin\theta_{l_2}-\sin\theta_{l_1}$,  $f(x_0)$ can be rewritten as
\begin{equation}
\begin{aligned}
f(\pi(\phi_{l_2}-\phi_{l_1}))
&=\left|\frac{\lambda}{N_t\pi d\Delta}\right|.
\end{aligned}
\end{equation}
When  $N_t\to \infty$, we have
\begin{equation}
\begin{aligned}
\lim\limits_{N \to \infty}f(\pi(\phi_{l_2}-\phi_{l_1}))
&=\lim\limits_{N \to \infty}\left|\frac{\lambda}{N_t\pi d\Delta}\right|= 0.
\end{aligned}
\end{equation}
At this time, the proof of {\bf Lemma 1} is completed.
\section{}
Since $h_l$ follows $\mathcal{CN}(0,1)$, the variance of both the real and imaginary parts is $1/2$.
In this respect, $|h_l|^2$ follows the central Chi-square distribution with two DoF.
After some mathematical operations, it can obtain
\begin{equation}\label{tradfF}
\ddot f(x) = \exp(-x), \ \ F(x) = 1- \exp(-x).
\end{equation}

After sorting the combinations as in the descending order,
the PDF of $|h_l|^2, l \in \{1,2,\cdots,L\}$, can be expressed as \cite{orderstat}
\begin{equation}\label{bino}
\begin{aligned}
{{f}}(x)&=\frac{1}{\mathcal{B}(L-l+1,l)}\left[{F}(x)\right]^{L-l}
\left[1-{F}(x)\right]^{l-1}{\ddot f}(x).
\end{aligned}
\end{equation}
Since $L-l+1$ and $l$ are both integers, beta function can be rewritten as \cite{jef2007book}
\begin{equation}\label{betaa}
\mathcal{B}(a,b)=\frac{\Gamma(a)\Gamma(b)}{\Gamma(a+b)}.
\end{equation}
Inserting Eqs. (\ref{tradfF}) and (\ref{betaa}) into Eq. (\ref{bino}), the PDF can be expressed as
\begin{equation}\label{bin1}
f(x)
=\frac{\Gamma(L+1)}{\Gamma(L-l+1)\Gamma (l)}\left[1-\exp\left(-{x}\right)\right]^{L-l}{[\exp\left(-{ x}\right)]^{l}}.
\end{equation}
By using binomial theorem \cite{jef2007book}, we can simplify the Eq. (\ref{bin1}) as
\begin{equation}\label{pdf11}
\begin{aligned}
f(x)
=&\frac{L!}{(L-l)!(l-1)!}\sum_{k=0}^{L-l}\dbinom{L-l}{k}\\&\times
\left[-\exp\left(-{x}\right)\right]^k{\exp\left(-{l x}\right)}.
\end{aligned}
\end{equation}
After some simple algebraic operations, we complete the proof of {\bf Lemma 2}.

\section{}
According to the definition of MGF, we have
\begin{equation}\label{mgf1}
\begin{aligned}
{\rm MGF}_x(s)
&=\int_0^\infty f(x)e^{sx}dx.
\end{aligned}
\end{equation}
Substituting  Eq. (\ref{bino2}) into  Eq. (\ref{mgf1}), we get
\begin{equation}\label{xsdf}
\begin{aligned}
{\rm MGF}_x(s)
&=\frac{L!}{(L-l)!(l-1)!}\int_0^\infty \left(1-e^{-{x}}\right)^{L-l}(e^{-x})^{(l-s)} dx.
\end{aligned}
\end{equation}
To facilitate the derivation, we let $t$ = $1-e^{-x}$. Here, the ${\rm MGF}_x(s)$ in  Eq. (\ref{xsdf}) can be updated as
\begin{equation}\label{xsdf1}
\begin{aligned}
{\rm MGF}_x(s)
&\overset{}{=}\frac{L!}{(L-l)!(L_s-1)!}\int_0^1 t^{(L-l)}(1-t)^{(l-s-1)}dt.
\end{aligned}
\end{equation}
It is observed that the integral term with respect to the variable $t$ can be expressed in terms of Beta function.
According to \cite{jef2007book}, the MGF in  Eq. (\ref{xsdf1}) can be re-expressed as
\begin{equation}\label{xsdf2}
{\rm MGF}_x(s)
\overset{}{=}\frac{L!}{(L-l)!(l-1)!}\mathcal{B}(L-l+1,l-s).
\end{equation}
By using  Eq. (\ref{betaa}), MGF in  Eq. (\ref{xsdf2}) can be evaluated as
\begin{equation}\label{xsdf3}
{\rm MGF}_x(s)=\frac{L!(l-s-1)!}{(l-1)!(L-s)!}.
\end{equation}
On the basis of this, we unfold the factorial function to obtain
\begin{equation}\label{MGFw1}
\begin{aligned}
{\rm MGF}_x(s)
&=\frac{L_s\times(l+1)\times\cdots\times L}{(l-s)\times(l+1-s)\times\cdots\times (L-s)}.
\end{aligned}
\end{equation}
Here, the proof of {\bf Lemma 3} is completed.
\section{}
For the case of $\hat l = l$,
the UPEP in  Eq. (\ref{ana11}) can be rewritten via the $Q(x) \approx \frac{1}{12}\exp\left(-\frac{x^2}{2}\right)+\frac{1}{4}\exp\left(-\frac{2x^2}{3}\right)$ as
\begin{equation}
\begin{aligned}
\bar{P}_e
=&\frac{1}{12}\int_0^\infty f(x)\exp\left(-\frac{\rho |s_m - s_{\hat m}|^2x}{4}\right)dx\\&+\frac{1}{4}\int_0^\infty f(x)\exp\left(-\frac{\rho |s_m - s_{\hat m}|^2x}{3}\right)dx\\
=&\frac{1}{12}\prod \limits_{\xi=l}^L\frac{4\xi}{4\xi+{\rho |s_m - s_{\hat m}|^2}}+\frac{1}{4}\prod \limits_{\xi=l}^L\frac{3\xi}{3\xi+{\rho |s_m - s_{\hat m}|^2}}\\
=&\frac{1}{12}\prod \limits_{\xi=l}^L\left(\frac{48\xi^2+13\rho |s_m - s_{\hat m}|^2\xi}{12\xi^2+7\rho |s_m - s_{\hat m}|^2\xi+({\rho |s_m - s_{\hat m}|^2})^2}\right).
\end{aligned}
\end{equation}
According to Eq. (\ref{upepee}), the proof of {\bf Theorem 1} is completed.

\section{}
The mutual information can be described as
\begin{equation}\label{xxxx}
\mathcal{I}(h_l,s_m;\mathbf{y}_r(l)) = \mathcal{I}(\mathbf{y}_r(l))-\mathcal{I}(\mathbf{y}_r(l)|h_l,s_m ),
\end{equation}
where $\mathcal{I}(\mathbf{y}_r(l))$ denotes the differential entropy between the received signals
and $\mathcal{I}(\mathbf{y}_r(l)|h_l,s_m )$ stands for the differential entropy of noise $n_r$.
The $\mathcal{I}(\mathbf{y}_r(l)|h_l,s_m )$ can be written as
\begin{equation}\label{h1q1}
\mathcal{I}(\mathbf{y}_r(l)|h_l,s_m ) = \log_2(\pi\exp(1)N_0).
\end{equation}
Based on \cite{bock2008mutu}, $\mathcal{I}(\mathbf{y}_r(l))$ can be characterized as
\begin{equation}
\mathcal{I}(\mathbf{y}_r(l))=-\int f(\mathbf{y}_r(l))\log_2(f(\mathbf{y}_r(l)))d\mathbf{y}_r(l).
\end{equation}
Next, the lower bound of $\mathcal{I}(\mathbf{y}_r(l))$ can be evaluated as \cite{guo2016on}
\begin{equation}\label{h1q2}
\mathcal{I}(\mathbf{y}_r(l))\geq -\log_2\left[ \int f^2(\mathbf{y}_r(l))\right]d\mathbf{y}_r(l).
\end{equation}
After some mathematical operations, the PDF of $\mathbf{y}_r(l)$ can be formulated as
\begin{equation}
\begin{aligned}
f(\mathbf{y}_r(l))&=
\frac{1}{L_sM}\sum_{l=1}^{L_s}\sum_{m=1}^M f(\mathbf{y}_r(l)|h_l,s_m)\\
&=\frac{1}{L_sM}\sum_{l=1}^{L_s}\sum_{m=1}^M \frac{1}{\pi N_0}\\&
\times\exp\left(
-\frac{|\mathbf{y}_{{r}}\left(l\right)-\boldsymbol{a}_r^H(\theta_{ l}^r)\mathbf{H}{\boldsymbol{a}_t(\theta^t)}\sqrt{P_s}s_m |^2}{N_0}
\right).
\end{aligned}
\end{equation}
In accordance with \cite{bock2008mutu}, we substitute Eqs. (\ref{h1q1}) and (\ref{h1q2}) into Eq. (\ref{xxxx}) and then adopt the Eq. (\ref{rorthh}).
At this time, we have
\begin{equation}\label{egrodi1}
\begin{aligned}
&\mathcal{I}(h_l,x_m;\mathbf{y}_r(l)) \approx 2\log_2(L_sM)-\log_2\left(
L_sM\right.\\&\left.+\sum_{l=1}^{L_s}\sum_{m=1}^M\sum_{\hat{l}\neq l =1}^{L_s} \sum_{\hat m \neq m =1}^{M}
\exp\left(-\frac{\sqrt{P_s}|h_{\hat l}s_{\hat{m}}|^2}{N_0}\right)
\right).
\end{aligned}
\end{equation}
Subsequently, we calculate the expectation of Eq. (\ref{egrodi1}) to obtain the ergodic capacity of the proposed scheme, which is denoted as
\begin{equation}\label{eco2}
\begin{aligned}
C&= 2\log_2(L_sM)-\log_2\left(
L_sM\right.\\&\left.+\sum_{l=1}^{L_s}\sum_{m=1}^M\sum_{\hat{l}\neq l =1}^{L_s} \sum_{\hat m \neq m =1}^{M}
\mathbb{E}\left[\exp\left(-\frac{\rho|h_{\hat l}s_{\hat{m}}|^2|}{2}\right)\right]
\right).
\end{aligned}
\end{equation}
According to the definition of MGF, we have
\begin{equation}\label{ecf2}
{\rm MGF}_{|h_{\hat l}|^2}\left( -\frac{\rho|h_{\hat l}s_{\hat{m}}|^2|}{2}\right) = \mathbb{E}\left[\exp\left(-\frac{\rho|s_{\hat{m}}|^2|}{2}\right)\right].
\end{equation}
By substituting Eq. (\ref{ecf2}) into Eq. (\ref{eco2}), the proof of {\bf Proposition 1} is completed.

\end{appendices}

\end{document}